\documentstyle[aps,prc,floats,graphicx,tighten]{revtex}

\begin{document}

\def\nuc#1#2{${}^{#1}$#2}
\def\mnuc#1#2{{}^{#1}{\protect\text{#2}}}
\def\E#1{$\! \times$\,10$^{#1}$}
\def\+/-{$\pm$}
\def\about{$\sim\,$}
\def\etal{{\it et al}.}
\def\degreesC{$\,^{\circ}$C}
\def\today{\space\number\day\space\ifcase\month\or
  January\or February\or March\or April\or May\or June\or July\or
  August\or September\or October\or November\or December\fi
  \space\number\year}

\draft

\twocolumn[\hsize\textwidth\columnwidth\hsize%
\csname@twocolumnfalse\endcsname

\begin{flushright}
hep-ph/9803418 \\
Phys.\ Rev.\ C {\bf 59}, 2246-2263 (April 1999)
\end{flushright}

\title{Measurement of the response of a gallium metal solar
neutrino experiment \\ to neutrinos from a $^{\bbox{51}}$Cr
source}

\author{J.\,N.\,Abdurashitov, V.\,N.\,Gavrin, S.\,V.\,Girin,
V.\,V.\,Gorbachev, T.\,V.\,Ibragimova, A.\,V.\,Kalikhov,
N.\,G.\,Khairnasov, T.\,V.\,Knodel,
V.\,N.\,Kornoukhov\footnotemark, I.\,N.\,Mirmov, A.\,A.\,Shikhin,
E.\,P.\,Veretenkin, V.\,M.\,Vermul, V.\,E.\,Yants, and
G.\,T.\,Zatsepin}
\address{Institute for Nuclear Research, Russian Academy of
Sciences, 117312 Moscow, Russia}

\author{Yu.\,S.\,Khomyakov and A.\,V.\,Zvonarev\footnotemark}
\address{Institute of Physics and Power Engineering, Obninsk,
Russia}

\author{T.\,J.\,Bowles, J.\,S.\,Nico\footnotemark,
W.\,A.\,Teasdale, and D.\,L.\,Wark\footnotemark}
\address{Los Alamos National Laboratory, Los Alamos, New Mexico
87545}

\author{M.\,L.\,Cherry}
\address{Louisiana State University, Baton Rouge, Louisiana
70803}

\author{V.\,N.\,Karaulov, V.\,L.\,Levitin, V.\,I.\,Maev,
P.\,I.\,Nazarenko, V.\,S.\,Shkol'nik, and N.\,V.\,Skorikov}
\address{Mangyshlak Atomic Energy Complex, Aktau, Kazakhstan}

\author{B.\,T.\,Cleveland, T.\,Daily, R.\,Davis, Jr.\,,
K.\,Lande, C.\,K.\,Lee, and P.\,S.\,Wildenhain}
\address{University of Pennsylvania, Philadelphia, Pennsylvania
19104}

\author{S.\,R.\,Elliott and J.\,F.\,Wilkerson}
\address{University of Washington, Seattle, Washington 98195}

\author{(The SAGE Collaboration)}

\date{Received 25 March 1998}
\maketitle

\begin{abstract}

     The neutrino capture rate measured by the Russian-American
Gallium Experiment is well below that predicted by solar models.
To check the response of this experiment to low-energy neutrinos,
a 517 kCi source of \nuc{51}{Cr} was produced by irradiating
512.7~g of 92.4\%-enriched \nuc{50}{Cr} in a high-flux fast
neutron reactor.  This source, which mainly emits monoenergetic
747-keV neutrinos, was placed at the center of a 13.1 tonne
target of liquid gallium and the cross section for the production
of \nuc{71}{Ge} by the inverse beta decay reaction
$\mnuc{71}{Ga}(\nu_e,e^-)\mnuc{71}{Ge}$ was measured to be
{[5.55 \+/-~0.60~(stat) \+/-~0.32~(syst)]} \E{-45} cm$^2$.  The
ratio of this cross section to the theoretical cross section of
Bahcall for this reaction is 0.95 \+/- 0.12 (expt)
$^{+0.035}_{-0.027}$ (theor) and to the cross section of Haxton
is 0.87 \+/- 0.11 (expt) \+/- 0.09 (theor).  This good agreement
between prediction and observation implies that the overall
experimental efficiency is correctly determined and provides
considerable evidence for the reliability of the solar neutrino
measurement.
\end{abstract}


] 

\footnotetext[1]{Present address: Institute of Theoretical and
Experimental Physics, 117259 Moscow, Russia.}
\footnotetext[2]{Deceased.}
\footnotetext[3]{Present address: National Institute of Standards
and Technology, Bldg. 235/A106, Gaithersburg, Maryland 20899.}
\footnotetext[4]{Present address: Department of Particle and
Nuclear Physics, Oxford University, Keble Road, Oxford OX1 3RH,
UK.}


\section{Introduction}

     Gallium experiments are uniquely able to measure the
principal component of the solar neutrino spectrum.  This is
because the low threshold of 233 keV \cite{Audi and Wapstra 95}
for inverse beta decay on the 40\% abundant isotope \nuc{71}{Ga}
is well below the end point energy of the neutrinos from
proton-proton fusion, which are predicted by standard solar
models to be about 90\% of the total flux.  The Russian-American
Gallium Experiment (SAGE) has been measuring the capture rate of
solar neutrinos with a target of gallium metal in the liquid
state since January 1990.  The measured capture rate
\cite{Abdurashitov et al. 94,Gavrin 98} is
67~\+/-7~(stat)~$^{+5}_{-6}$ (syst) SNU\footnote{1 SNU
corresponds to one neutrino capture per second in a target that
contains 10$^{36}$ atoms of the neutrino absorbing isotope.}, a
value that is well below solar model predictions of 137
$^{+8}_{-7}$ SNU \cite{Bahcall and Pinsonneault and Wasserburg
95} and 125~\+/- 5 SNU \cite{Turck-Chieze and Lopes 93}.  In
addition, the GALLEX Collaboration, which has been measuring the
solar neutrino capture rate with an aqueous GaCl$_3$ target since
1991, observes a rate of 70~\+/- 7~$^{+4}_{-5}$ SNU \cite{Hampel
et al. 96}.

     The other two operating solar neutrino experiments, the
chlorine experiment \cite{Cleveland et al. 98} and the Kamiokande
experiment \cite{Suzuki et al. 95}, have significantly
higher-energy thresholds, and thus are not able to see the
neutrinos from $pp$ fusion.  When the results of these four solar
neutrino experiments are considered together, a contradiction
arises which cannot be accommodated by current solar models, but
which can be explained if one assumes that neutrinos can
transform from one species to another \cite{Bahcall 94,Berezinsky
94,Parke 95,Hata et al. 94,Castellani et al. 94,Bahcall et al.
95,Heeger and Robertson 96}.

     The gallium experiment, in common with other radiochemical
solar neutrino experiments, relies on the ability to extract,
purify, and count, all with well known efficiencies, a few atoms
of a radioactive element that were produced by neutrino
interactions inside many tonnes of the target material.  In the
case of 60 tonnes of Ga, this represents the removal of a few
tens of atoms of \nuc{71}{Ge} from $5 \times 10^{29}$ atoms of
Ga.  To measure the efficiency of extraction, about 700 $\mu$g of
stable Ge carrier is added to the Ga at the beginning of each
exposure, but even after this addition, the separation factor of
Ge from Ga is still 1 atom in 10$^{11}$.  This impressively
stringent requirement raises legitimate questions about how well
the many efficiencies that are factored into the final result are
known.  It has been understood since the outset that a rigorous
check of the entire operation of the detector (i.e., the chemical
extraction efficiency, the counting efficiency, and the analysis
technique) would be made if it is exposed to a known flux of
low-energy neutrinos.  In addition to verifying the operation of
the detector, such a test also eliminates any significant
concerns regarding the possibility that atoms of \nuc{71}{Ge}
produced by inverse beta decay may be chemically bound to the
gallium (so-called ``hot atom chemistry'') in a manner that
yields a different extraction efficiency than that of the natural
Ge carrier.  In other words, it tests a fundamental assumption in
radiochemical experiments that the extraction efficiency of atoms
produced by neutrino interactions is the same as that of carrier
atoms.

     This article describes such a test, in which a portion of
the SAGE gallium target was exposed to a known flux of
\nuc{51}{Cr} neutrinos and the production rate of \nuc{71}{Ge}
was measured.  Similar tests have also been made by GALLEX
\cite{Hampel et al. 98}.

     Although a direct test with a well-characterized neutrino
source lends significant credibility to the radiochemical
technique, we note that numerous investigations have been
undertaken during the SAGE experiment to ensure that the various
efficiencies are as quoted \cite{Abdurashitov et al. 94}.  The
extraction efficiency has been determined by a variety of
chemical and volumetric measurements that rely on the
introduction and subsequent extraction of a known amount of the
stable Ge carrier.  A test was also carried out in which Ge
carrier doped with a known number of \nuc{71}{Ge} atoms was added
to 7 tonnes of Ga.  Three standard extractions were performed,
and it was demonstrated that the extraction efficiencies of the
carrier and \nuc{71}{Ge} follow each other very closely.

     Another experiment was performed to specifically test the
possibility that atomic excitations might tie up \nuc{71}{Ge} in
a chemical form from which it would not be efficiently extracted.
There is a concern that this might occur in liquid gallium
because the metastable Ga$_2$ molecule exists with a binding
energy of \about1.6 eV.  In this experiment the radioactive
isotopes \nuc{70}{Ga} and \nuc{72}{Ga} were produced in liquid
gallium by neutron irradiation.  These isotopes quickly beta
decay to \nuc{70}{Ge} and \nuc{72}{Ge}.  The Ge isotopes were
extracted from the Ga using our standard procedure and their
number was measured by mass spectrometry.  The results were
consistent with the number expected to be produced based on the
known neutron flux and capture cross section, thus suggesting
that chemical traps are not present.  This experiment is not
conclusive, however, because the maximal energy imparted to the
\nuc{70}{Ge} nucleus following beta decay of \nuc{70}{Ga} is 32
eV, somewhat higher than the maximal energy of 20 eV received by
the \nuc{71}{Ge} nucleus following capture of a 747-keV neutrino
from \nuc{51}{Cr} decay (and considerably higher than the maximum
nuclear recoil energy of 6.1 eV after capture of a 420-keV
neutrino from proton-proton fusion).

     Further evidence that the extraction efficiency was well
understood came from monitoring the initial removal from the Ga
of cosmogenically produced \nuc{68}{Ge}.  This nuclide was
generated in the Ga as it resided on the surface exposed to
cosmic rays.  When the Ga was brought underground, the reduction
in the \nuc{68}{Ge} content in the initial extractions was the
same as for the Ge carrier.  These numerous checks and auxiliary
measurements have been a source of confidence in our methodology,
yet it is clear that a test with an artificial neutrino source of
known activity provides the most compelling validation of
radiochemical procedures.

     This article is an elaboration of work that previously
appeared in Ref.~\cite{Abdurashitov et al. 96}.  The experimental
changes since the previous Letter are some minor refinements in
the selection of candidate \nuc{71}{Ge} events and in the
treatment of systematic errors; recent cross section calculations
are also included.  The central experimental result given here is
almost identical to what was reported earlier.

\section{$^{\bf51}$C\lowercase{r} Source}
\subsection{Choice of chromium}

     A number of K-capture isotopes that can be produced by
neutron irradiation in a high-flux reactor, \nuc{51}{Cr},
$^{65}$Zn, and \nuc{37}{Ar}, have been suggested \cite{Kuzmin
67,Alvarez 73,Raghavan 78,Haxton 88} as sources that can be used
to check the response of solar neutrino detectors.  The isotope
\nuc{51}{Cr} emits neutrinos with energy closest to the $pp$
neutrinos, the solar neutrino component to which gallium is most
sensitive, and thus is the best choice for the gallium
experiments.

\begin{figure}[ht]
\begin{center}
\includegraphics*[width=3.375in,bb=103 89 364 241] {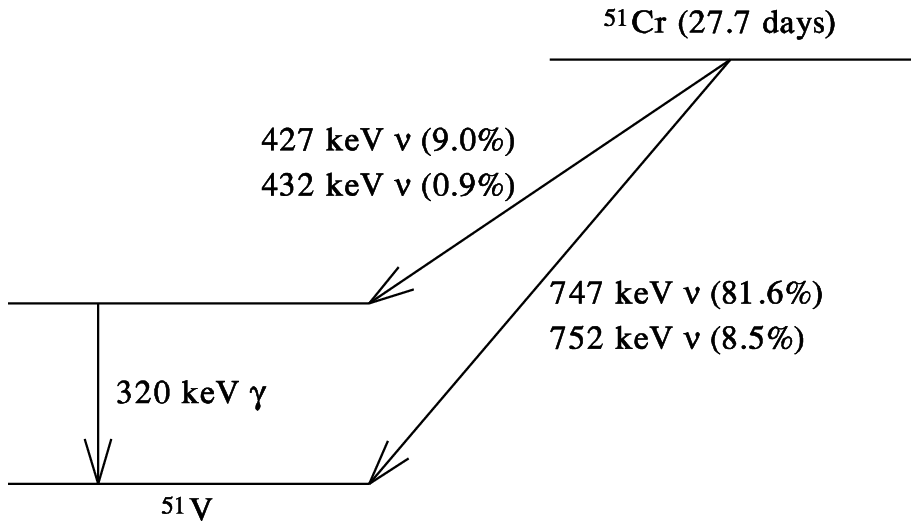}
\caption{Decay scheme of \nuc{51}{Cr} to \nuc{51}{V} through
electron capture.}
\label{decay scheme}
\end{center}
\end{figure}

     The decay of \nuc{51}{Cr} is by electron capture,
$\mnuc{51}{Cr} + e^- \rightarrow \mnuc{51}{V} + \nu_e$, with a
half-life of 27.7 d.  The decay scheme is illustrated in
Fig.\ \ref{decay scheme}.  There is a 90.12\% branch\cite{Table
of Isotopes 96} that decays directly to the ground state of
\nuc{51}{V} and a 9.88\% branch which decays to the first excited
state of \nuc{51}{V}, which promptly decays with the emission of
a 320-keV gamma ray to the ground state.  Taking into account the
atomic levels to which transitions can occur, the neutrino
energies are 752 keV (9\%), 747 keV (81\%), 432 keV (1\%), and
427 keV (9\%).

     The intensity of the \nuc{51}{Cr} source must be high enough
that the production rate in the gallium target is significantly
greater than the solar neutrino capture rate.  The source
activity is thus required to be near to 1 MCi, far surpassing the
activity of most sources produced at reactors.  Because
\nuc{50}{Cr} has only a 4.35\% natural abundance, it is
impossible to produce the necessary activity of \nuc{51}{Cr} by
irradiation of natural Cr in any presently existing nuclear
reactor.  The required activity can only be attained by
irradiation of enriched \nuc{50}{Cr}, as shown in
Ref.~\cite{Gavrin et al. 84} and additionally considered in
Ref.~\cite{Cribier et al. 88}.  Besides decreasing the irradiated
mass to a value that can be acceptably placed in a reactor, the
use of enriched Cr reduces the self-shielding during irradiation
and reduces the neutron competition from \nuc{53}{Cr}, whose
capture cross section for thermal neutrons is very high.

\begin{table}[ht]
\narrowtext
\caption{Isotopic composition of natural Cr and of the enriched
Cr in the source.}
\label{Cr isotopic composition}
\begin{tabular}{c d c}
          & \multicolumn{2}{c}{Abundance (\%)}              \\
\cline{2-3}
Isotope   &        Natural     &  Enriched               \\
\hline
50        &              4.35  &        92.4 \+/- 0.5    \\
52        &             83.79  &         7.6 \+/- 0.4    \\
53        &              9.50  &      $<$0.5             \\
54        &              2.36  &      $<$0.2
\end{tabular}
\end{table}

     The chromium used in our experiment was enriched to 92.4\%
in \nuc{50}{Cr}.  The isotopic composition is given in
Table~\ref{Cr isotopic composition}.  The advantage of this high
enrichment was that it yielded a source of great specific
activity (more than 1 kCi/g) and small physical size, thus giving
a very high neutrino capture rate.

\subsection{Cr preparation}

     Enriched chromium was produced by the Kurchatov Institute by
gas centrifugation of chromium oxyfluoride, CrO$_2$F$_2$
\cite{Tikhomorov 92,Popov et al. 95}.  The highly corrosive
CrO$_2$F$_2$ was then hydrolyzed to chromium oxide, Cr$_2$O$_3$.
To obtain an extremely compact source, the chromium oxide was
then reduced to metallic chromium.  This reduction was done by
heating a cold-pressed mixture of chromium oxide and high purity
graphite in a hydrogen atmosphere; the resulting product was
melted in an Al$_2$O$_3$ crucible to remove gaseous impurities.
The Cr ingots were then crushed into pieces of 1--3 mm size and
the chromium was treated with hydrogen at 1200\degreesC for 24 h
to remove residual oxygen and nitrogen.

     For the reactor irradiation the metallic Cr was extruded
into the form of rods, 45 mm long by 7 mm in diameter.  Cr chips
were placed into a molybdenum-lined steel shell under modest
pressure at room temperature and the shell was electron beam
welded in vacuum (10$^{-6}$ torr).  This shell with the enclosed
chromium was subjected to very high pressure at 1100\degreesC for
30 seconds and then extruded at 1000\degreesC such that the
length of the Cr was increased by a factor of 7.  The steel shell
was then dissolved in nitric acid and chromium rods of the
desired size were produced by machining and spark cutting.
Finally, the rods were recrystallized at 1000\degreesC.  The
measured density (taking into account the Cr isotopic
composition), grain size, and hardness of the resulting rods were
very close to those of pure defectless metallic chromium.  Table
\ref{Cr rod properties} gives the properties of the 50 rods that
were prepared.  Their microstructure was essentially identical to
that of pure metallic Cr.

\begin{table}[ht]
\narrowtext
\caption{Properties of the Cr rods prepared for irradiation.}
\label{Cr rod properties}
\begin{tabular}{ l d d d }
Characteristic         &   Lot 1    &   Lot 2    &  Lot 3    \\
\hline
Number of rods         &      21.   &      21.   &      8.   \\
Total mass of Cr (g)   &     245.8  &     244.9  &     93.43 \\
Hardness (kg/mm$^2$)   &     138.   &     134.   &    152.   \\
Grain size ($\mu$m)    &      18.   &      24.   &     23.   \\
Density (g/cm$^3$)     &       6.93 &       6.96 &      6.96
\end{tabular}
\end{table}

\subsection{Cr irradiation and source assembly}

     The Cr was irradiated at the BN-350 fast breeder nuclear
reactor at the atomic power station in Aktau, Kazakhstan.  This
reactor was designed for simultaneous power and secondary nuclear
fuel production.  Other similar reactors are BN-600 in Russia,
Phenix and Super Phenix in France, and MONJU in Japan.  BN-350
has a core of highly enriched uranium without a moderator and a
blanket of unenriched uranium; liquid Na is used as a coolant.
This construction gives a high flux of fast neutrons [to
5 \E{15}/(cm$^2$ s)] at nominal power, which is advantageous for
making intense sources \cite{Zvonarev et al. 96}.

\begin{figure}[ht]
\begin{center}
\includegraphics*[height=4.375in,bb=182 38 335 747]
{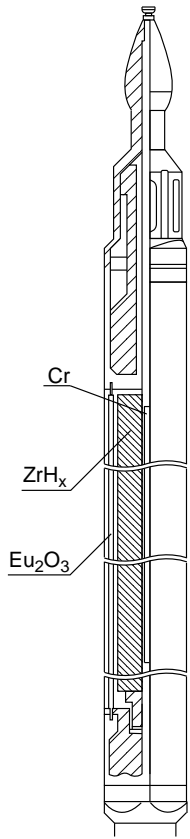}
\caption{Irradiation assembly (IA).}
\label{irradiation assembly}
\end{center}
\end{figure}

\begin{figure}[ht]
\begin{center}
\includegraphics*[width=2.67in,bb=19 16 229 271] {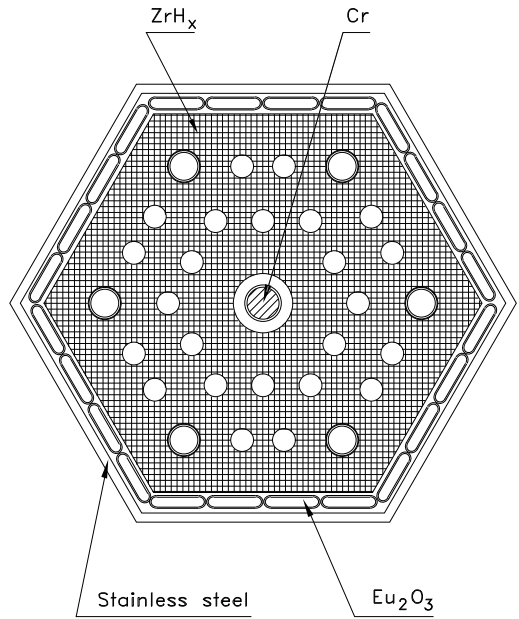}
\caption{Cross section of irradiation assembly (IA).  The open
circles are cooling channels for liquid Na.}
\label{irradiation assembly cross section}
\end{center}
\end{figure}

     The cross section for capture of fast neutrons by
\nuc{50}{Cr} is less than 0.1 b, much too low to reach the
desired specific activity of \nuc{51}{Cr}.  Therefore a unique
irradiation assembly (IA) was developed, which could be placed in
the BN-350 core as a replacement fuel assembly (see
Fig.\ \ref{irradiation assembly} and Fig.\ \ref{irradiation
assembly cross section}).  Most of the volume of the IA consisted
of a zirconium hydride moderator around a central stainless steel
tube that contained the \nuc{50}{Cr} metal rods.  This gave a
high flux of low-energy neutrons in the vicinity of the
\nuc{50}{Cr}, and as a result, a much increased average capture
cross section (\about4 b).  To prevent leakage of these low
energy neutrons from the IA, which could increase the power
release in neighboring fuel assemblies, the moderator was
surrounded by absorbing elements made of europium oxide.
Finally, the presence of the IA results in a negative reactivity
effect.  To compensate for this the standard configuration of the
reactor core was altered by replacing a few assemblies with ones
of higher fuel enrichment and by installing a few additional fuel
assemblies.

\begin{figure}[ht]
\begin{center}
\includegraphics*[height=3.375in,bb=165 75 459 644]
{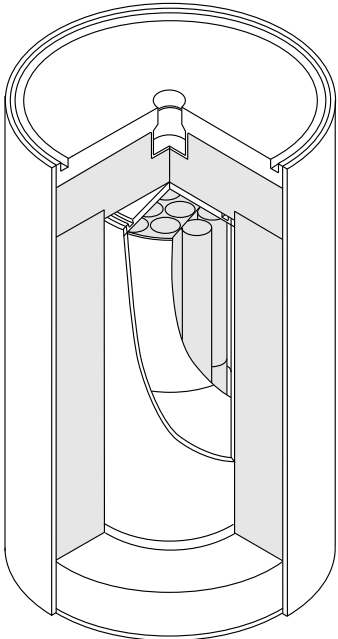}
\caption{Cut-away drawing of the source.  The Cr rods were placed
within the inner cylinders.}
\label{drawing of source}
\end{center}
\end{figure}

     Calculations showed that by using two irradiation assemblies
we would expect to produce a source whose activity at the end of
irradiation was between 0.5 and 0.8 MCi, depending on reactor
power and the position in the reactor core where the IA was
irradiated.  The final physical characteristics of the IA were
measured in a low-power experiment, which was carried out in the
BN-350 reactor before the full-scale irradiation.  This
experiment showed safe irradiation of the IA but gave less
\nuc{51}{Cr} activity than anticipated.  To compensate for this
reduced activity it was decided to increase the reactor power
near the end of irradiation.  The IA's were installed on 4
September 1994 with the reactor power set at its usual level of
520 MW.  Irradiation continued until 2 December, at which time
the power was increased to 620 MW, so as to increase the final
\nuc{51}{Cr} activity.  The IA's were removed from the reactor on
18 December 1994.  Using remote manipulators inside a hot cell,
all 46 irradiated Cr rods were removed from the IA's and 44 of
them, whose total mass before irradiation was 512.7~g, were
placed into holes in a tungsten holder.  This holder was then put
into a stainless steel casing and the assembly was welded shut
and leak checked in a helium atmosphere.  This source assembly
was placed in a specially constructed tungsten radiation shield
with 18 mm wall thickness, which had an outer stainless steel
casing of 80 mm diameter by 140 mm height.  The outer stainless
shell was also welded shut and leak checked.  A cut away view of
the overall source assembly is shown in
Fig.\ \ref{drawing of source}.  This source was placed into a
shipping cask, flown to the Mineralnye Vody airport in southern
Russia, and then transported by truck to the Baksan Neutrino
Observatory where the \nuc{51}{Cr} irradiations of the gallium
were carried out.

\subsection{Source impurities}
\label{source impurities}

     There exist a large number of chemical elements that, upon
irradiation, produce long-lived gamma-emitting isotopes.  The
presence of these gamma emitters in the source must be strictly
controlled because they increase the size of the source shield
necessary for personnel protection and thus decrease the
effective neutrino path length in the gallium target, and they
add heat to the source and thus confuse the calorimetric
measurement of source activity which will be described below.
Because of their high capture cross section for thermal neutrons,
even minute quantities of some elements cannot be tolerated.

\begin{table*}[ht]
\caption{Measured impurities in the Cr rods prior to activation
and the predicted resulting activities at the reference time
(18:00 on 26 Dec.\ 1994).  These are compared to the measured
activities.}
\label{impurities}
\begin{tabular}{l d c d d c}
         &      Measured   &               &           &
\multicolumn{2}{c}{Activity (mCi)}\\ \cline{5-6}
Impurity & Content (ppm)   &    Nuclide    & Half-life &              Expected
&     Measured        \\
\hline
Fe       &            50.0 &  \nuc{59}{Fe} &   44.5 d  &                 9.
&    24 \+/- 3        \\
W        &            25.0 &  \nuc{187}{W} &   23.9 h  &                23.
&    negligible       \\
Cu       &            15.0 &  \nuc{64}{Cu} &   12.7 h  &               $<$0.1
&    negligible       \\
Ga       &             5.7 &  \nuc{72}{Ga} &   14.1 h  &               $<$0.1
&    negligible       \\
Na       &             3.3 &  \nuc{24}{Na} &   15.0 h  &               $<$0.1
&    negligible       \\
Zn       &             3.3 &  \nuc{65}{Zn} &   244. d  &                17.
&    negligible       \\
Ta       &             3.0 & \nuc{182}{Ta} &   115. d  &               1930.
&    38 \+/- 5        \\
Co       &             1.0 &  \nuc{60}{Co} &   5.3  y  &                81.
&    65 \+/- 15       \\
Sc       &             0.9 &  \nuc{46}{Sc} &   83.3 d  &                860.
&  1400 \+/- 100      \\
As       &             0.6 &  \nuc{76}{As} &   26.3 h  &                 2.
&    negligible       \\
Sb       &          $<$0.1 & \nuc{124}{Sb} &   60.2 d  &               $<$13.
&    negligible       \\
La       &          $<$0.1 & \nuc{140}{La} &   40.3 h  &               $<$0.4
&    negligible         
\end{tabular}
\end{table*}

     As a consequence, special care was taken in all the stages
of chemical processing to minimize contamination of the Cr.  To
be confident that the final impurity content of the Cr rods was
satisfactory, each rod was chemically analyzed before irradiation
by ICP-mass spectrometry with laser ablation and by spark mass
spectrometry.  The concentrations of the most relevant impurities
are given in Table \ref{impurities}, together with the expected
and measured activities after irradiation.  The dose rate at the
side surface of the source was 1.7 Sv/h at the beginning of the
first exposure (26 Dec.\ 1994) and only \about2\% of this was due
to impurities (mostly \nuc{46}{Sc}).  At the end of the last
exposure (23 May 1995), the dose rate had decreased to 0.05 Sv/h,
with an increase in the fraction due to impurities to \about25\%.

\begin{figure}[ht]
\begin{center}
\includegraphics*[width=3.375in,bb=27 13 437 286] {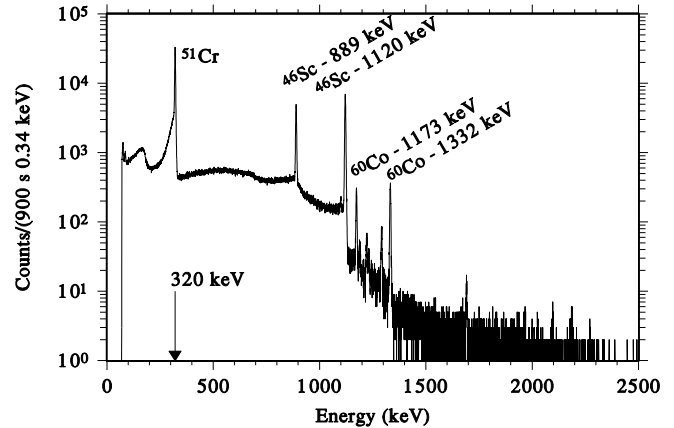}
\caption{Unshielded Ge detector spectrum of the gamma rays
emitted by the Cr source taken on 7 January 1995 at 10:40.  Gamma
lines are labeled by the isotope of origin.  Other contaminants
whose lines are not labeled include \nuc{59}{Fe}, \nuc{182}{Ta},
and \nuc{124}{Sb}.}
\label{Ge spectrum}
\end{center}
\end{figure}

\begin{table*}[ht]
\caption{Measured nuclide impurities in the \nuc{51}{Cr} source
and their contribution to the source activity measurement at the
reference time (18:00 on 26 Dec.\ 1994).  The power estimation
assumes that all the available energy is deposited in the
calorimeter.  A row with the data for the \nuc{51}{Cr} is
included for comparison.  The conversion constant 36.671
keV/decay was used in estimating the power for the Cr.}
\label{nuclide impurities}
\begin{tabular}{l d c d d}
               & Q value &     Measured     &   Estimated   &
Equivalent \nuc{51}{Cr}  \\
Isotope        &  (MeV)  &  activity (Ci)   &   power (W)   &
activity (kCi) \\
\hline
\nuc{46}{Sc}   &   2.37  &  1.400 \+/- 0.1  &     0.0200    &
0.092          \\
\nuc{60}{Co}   &   2.82  & 0.065 \+/- 0.015 &     0.0011    &
0.005          \\
\nuc{182}{Ta}  &   1.81  & 0.038 \+/- 0.005 &     0.0004    &
0.002          \\
\nuc{59}{Fe}   &   1.56  & 0.024 \+/- 0.003 &     0.0002    &
0.001          \\
Impurity total &         &                  &     0.0217    &        0.1         
\\
\\
\nuc{51}{Cr}   &   0.32  & 516600 \+/- 6000 &    112.3000   &
516.6
\end{tabular}
\end{table*}

     Figure~\ref{Ge spectrum} shows a gamma spectrum of the
source taken shortly after the start of the first Ga irradiation.
The 320-keV gamma ray from \nuc{51}{Cr} decay was attenuated by a
large factor by the tungsten shield, but still was the most
intense line in the spectrum.  The higher energy lines of
\nuc{46}{Sc}, \nuc{59}{Fe}, \nuc{60}{Co}, and \nuc{182}{Ta} had
much smaller attenuations and thus produced lines, even though
their activity was much lower than that of \nuc{51}{Cr}.  Limits
on the level of contamination activity can be inferred from this
spectrum and are summarized in Table \ref{nuclide impurities}.
The 1.5 Ci activity of \nuc{46}{Sc} was the largest single
contribution and the total activity of all contaminants was
estimated to be less than 2 Ci at this time.

     Table \ref{impurities} compares the values of activities
expected from the preirradiation impurity determinations and
those measured afterwards.  The only significant difference was
for Ta, which was because the mass-spectrometric analysis
preferentially sampled the surface of the Cr rods and not the
bulk material.  The apparently high concentration of Ta in the Cr
resulted from surface contamination by the tungsten carbide tool
used to machine the rods to the desired diameter.

\section{Extraction Schedule}

\begin{figure}[ht]
\begin{center}
\includegraphics*[width=3.375in,bb=154 78 425 188] {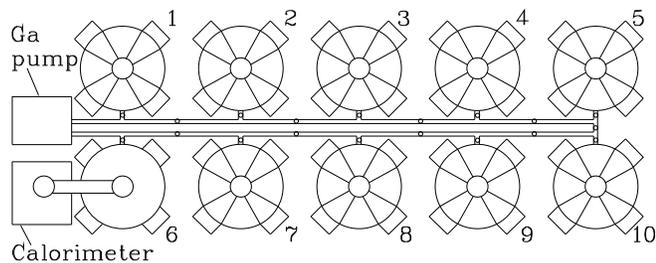}
\caption{Plan view of the laboratory showing the ten chemical
reactors, irradiation reactor 6 with the adjacent calorimeter,
and the Ga pump for transferring Ga between reactors.}
\label{reactor layout}
\end{center}
\end{figure}

\begin{figure}[t]
\begin{center}
\includegraphics*[width=3.375in,bb=0 -4 402 678] {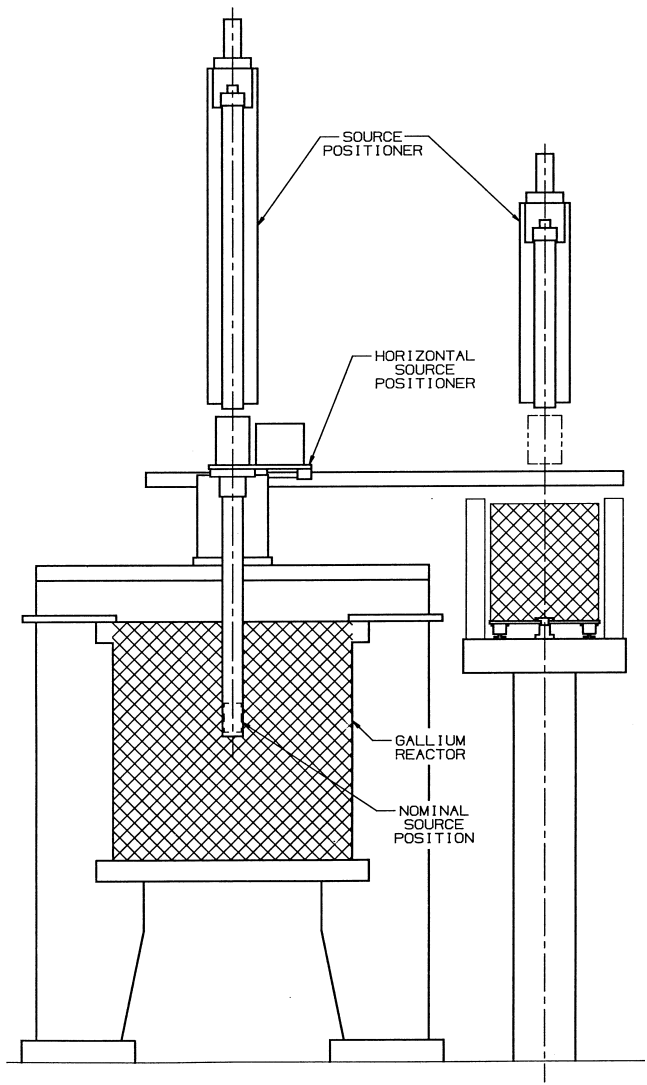}
\caption{Schematic drawing of the remote handling system which
moved the \nuc{51}{Cr} source from the gallium-containing reactor
to the adjacent calorimeter.}
\label{irradiation reactor}
\end{center}
\end{figure}

     The 55 tonnes of Ga that SAGE uses for solar neutrino
measurements is contained in eight chemical reactors with
approximately 7 tonnes in each.  Figure~\ref{reactor layout}
shows the layout of the ten reactors in the experimental area and
gives their numerical identification assignments.  In normal
solar neutrino operation Ga is contained in reactors 2--5 and
7--10.  All reactors except No.\ 6 are equipped with the
necessary mechanical equipment for the extraction process.
Reactor 6 was modified for the Cr exposures by removing its
stirring mechanism and replacing it with a reentrant Zr tube on
its axis which extended to the reactor center.  This modification
increased the capacity of reactor No.\ 6 to 13 tonnes of Ga.  To
begin each irradiation, a specially designed remote handling
system was used (Fig.\ \ref{irradiation reactor}) to place the
\nuc{51}{Cr} source inside this reentrant tube at the reactor
center.  At the end of each irradiation, the source was moved to
an adjacent calorimeter for activity measurement, and the gallium
was pumped back to the two reactors where it was stored during
solar neutrino runs.  The \nuc{71}{Ge} was then extracted with
the usual chemical procedures \cite{Gavrin et al. 89,Gavrin et
al. 94}.

     The source arrived at Mineralnye Vody on 20 Dec.\ 1994.
Because of a delay in customs approval, it was not delivered to
the laboratory in Baksan until 26 Dec.\ 1994.  The initial
installation of the source into the Ga was at 18:00 on 26 Dec.
We normalize all our results to this time.  Eight extractions
were conducted between 2 Jan.\ and 24 May 1995.  See Table
\ref{extraction data} for a summary of the extraction dates.  The
lengths of the exposure periods for the first five measurements
were chosen so each would have approximately equal statistical
uncertainty.  After these initial extractions, the Cr source had
decayed to the point where this was no longer possible and the
final three extractions were done at approximately monthly
intervals, the same schedule as for solar neutrino extractions.

\begin{table*}[ht]
\caption{Extraction schedule and related parameters.  The times
of exposure are given in days of year 1995.}
\label{extraction data}
\begin{tabular}{l c d d d d d d d d d}
Extraction     &Extraction          &   \multicolumn{2}{c}{Source
exposure}&\multicolumn{2}{c}{Source activity (kCi)} &
\multicolumn{2}{c}{Solar neutrino exposure}         &              Mass Ga
& \multicolumn{2}{c}{Extraction efficiency} \\ \cline{3-4}
\cline{5-6} \cline{7-8} \cline{10-11}
name     &date (1995)         &   Begin   &        End              &
Begin    &    End   &       Begin   &    End        &            (tonnes)
&     from Ga  &into GeH$_4$  \\
\hline
Cr 1     &   2 Jan. &      $-$4.25  &    1.86       &             516.6
&      443.4   & $-$10.58     &    2.05   &       13.123            &
0.85     &    0.82  \\
Cr 2     &   9 Jan. &        2.60   &    9.33       &             435.2
&      367.8   &   1.55       &    9.50   &       13.108            &
0.88     &    0.84  \\
Cr 2-2   &  11 Jan. &        2.60   &    9.33       &             435.2
&      367.8   &   9.50       &   11.46   &       13.094            &
0.85     &    0.71  \\
Cr 3     &  18 Jan. &        9.65   &   18.32       &             364.9
&      293.7   &   5.45       &   18.50   &       13.134            &
0.86     &    0.80  \\
Cr 4     &   3 Feb. &       19.04   &   34.32       &             288.5
&      196.8   &  18.50       &   34.49   &       13.119            &
0.89     &    0.85  \\
Cr 4-2   &   5 Feb. &       19.04   &   34.32       &             288.5
&      196.8   &  34.49       &   36.48   &       13.106            &
0.86     &    0.80  \\
Cr 5     &   1 Mar. &       34.84   &   60.46       &             194.3
&      102.3   &  11.46       &   60.63   &       13.081            &
0.90     &    0.83  \\
Cr 6     &  24 Mar. &       61.33   &   83.40       &             100.1
&       57.6   &  60.63       &   83.60   &       13.067            &
0.86     &    0.82  \\
Cr 6-2   &  26 Mar. &       61.33   &   83.40       &             100.1
&       57.6   &  83.60       &   85.45   &       13.054            &
0.85     &    0.79  \\
Cr 7     &  23 Apr. &       83.99   &   113.33      &              56.8
&       27.3   &  36.48       &   113.52  &       13.090            &
0.84     &    0.81  \\
Cr 7-2   &  26 Apr. &       83.99   &   113.33      &              56.8
&       27.3   &  113.52      &   116.46  &       13.077            &
0.87     &    0.72  \\
Cr 8     &   24 May &       118.83  &   143.86      &              23.8
&       12.7   &  116.46      &   144.54  &       13.063            &
0.92     &    0.82
\end{tabular}
\end{table*}

     The Cr experiment used reactors 6--10, shown in
Fig.\ \ref{reactor layout}.  To start the first exposure Ga was
pumped from reactors 9 and 10 to irradiation reactor 6 and then
the \nuc{51}{Cr} source was inserted to the center of this
reactor.  At the end of exposure 1 the source was moved to the
calorimeter and the Ga was pumped to reactors 9 and 10 for
extraction.  Immediately following this extraction, the Ga was
pumped to reactor 6 from reactors 9 and 10 and the source was
again placed at the center of reactor 6 to begin exposure 2.
Upon completion of exposure 2, the Ga was once again pumped to
reactors 9 and 10.  This time, however, two extractions were done
-- Nos.\ 2 and 2-2.  Meanwhile 13.134 tonnes of Ga from reactors
7 and 8 was pumped to reactor 6 to begin exposure 3.  This
pattern of exposure and extraction was repeated for a total of
eight exposures.  For exposures 1, 2, 5, and 6 the Ga was
extracted in reactors 9 and 10.  For exposures 3, 4, 7, and 8 the
Ga was extracted in reactors 7 and 8.  Second extractions
followed exposures 2, 4, 6, and 7.

     This extraction procedure differed somewhat from that used
for the solar neutrino experiment because there was the
additional step of the Ga transfer from two reactors to the
irradiation vessel and back.  Although there is no obvious reason
why this should introduce a change in extraction efficiency, a
number of tests were conducted to be confident that this
efficiency was not altered by the Ga transfer.  Prior to the
\nuc{51}{Cr} source exposure, nine solar neutrino extractions
were done from one or two reactors using all steps of the above
procedure including the Ga transfer.  The measured production
rate in these experiments was 92 SNU with a 68\% confidence range
from 53 SNU to 143 SNU.  This capture rate was entirely
consistent with that from solar neutrinos and no change was
observed in the counter background.

\section{Source Activity Determination}
\label{source activity determination}

\subsection{Source activity from calorimetry}

     The decay of \nuc{51}{Cr} deposits energy in the form of
heat in its surroundings.  Since all but 1 part in 10$^5$ of the
\nuc{51}{Cr} radiation was absorbed in the source, the source
activity could be determined by measuring its heat with a
calorimeter.  Table \ref{power input data} gives a summary of the
energy released in \nuc{51}{Cr} decay neglecting the energy lost
to neutrinos.  The average energy released which can be detected
as heat is 36.67~\+/- 0.20 keV/decay where the uncertainties have
been added in quadrature.

\begin{table*}[ht]
\caption{Summary of the input data to the power generated during
the decay of \nuc{51}{Cr}.  The value for the $M$-shell fraction
is deduced from the average of the $M/L$ ratios for the electron
capture isotopes \nuc{37}{Ar} and \nuc{55}{Fe}.}
\label{power input data}
\begin{tabular}{c @{\hspace{-2.0em}} d d d}
Type of radiation &                   Energy (keV)                  &
Fraction of \nuc{51}{Cr} decays            &               Energy released per
\nuc{51}{Cr} decay (keV)                   \\
\hline
Gamma             &   320.0852(9) \cite{Table of Isotopes 96,Zhou
91}               &      0.0988(5) \cite{Table of Isotopes 96}      &
31.624(160)       \\
$K$ capture       &               5.465 \cite{HCP 93}               &
0.895(5) \cite{Heuer 66}                   &                    4.891(25)
\\
$L$ capture       &               0.628 \cite{HCP 93}               &
0.0925(50) \cite{Heuer 66}                 &                     0.058(2)
\\
$M$ capture       &               0.067 \cite{HCP 93}               &
0.0125 (calc)     &                  0.001 (small)                  \\
Int.\,bremss.     &                 751 (end point)                 &
3.8 \E{-4} $\times$ 0.902 (\+/- \about 10\%)                        &
0.096(10) \cite{Table of Radioactive Isotopes 86}                   \\
Int.\,bremss.     &                 430 (end point)                 &
1.2 \E{-4} $\times$ 0.0983 (\+/- \about 10\%)                       &
0.001 (small)     \\
\\
Total             &                                                 &
                  &                   36.671(197)                    
\end{tabular}
\end{table*}

\begin{figure}[ht]
\begin{center}
\includegraphics*[width=2.86in,bb=178 255 477 642] {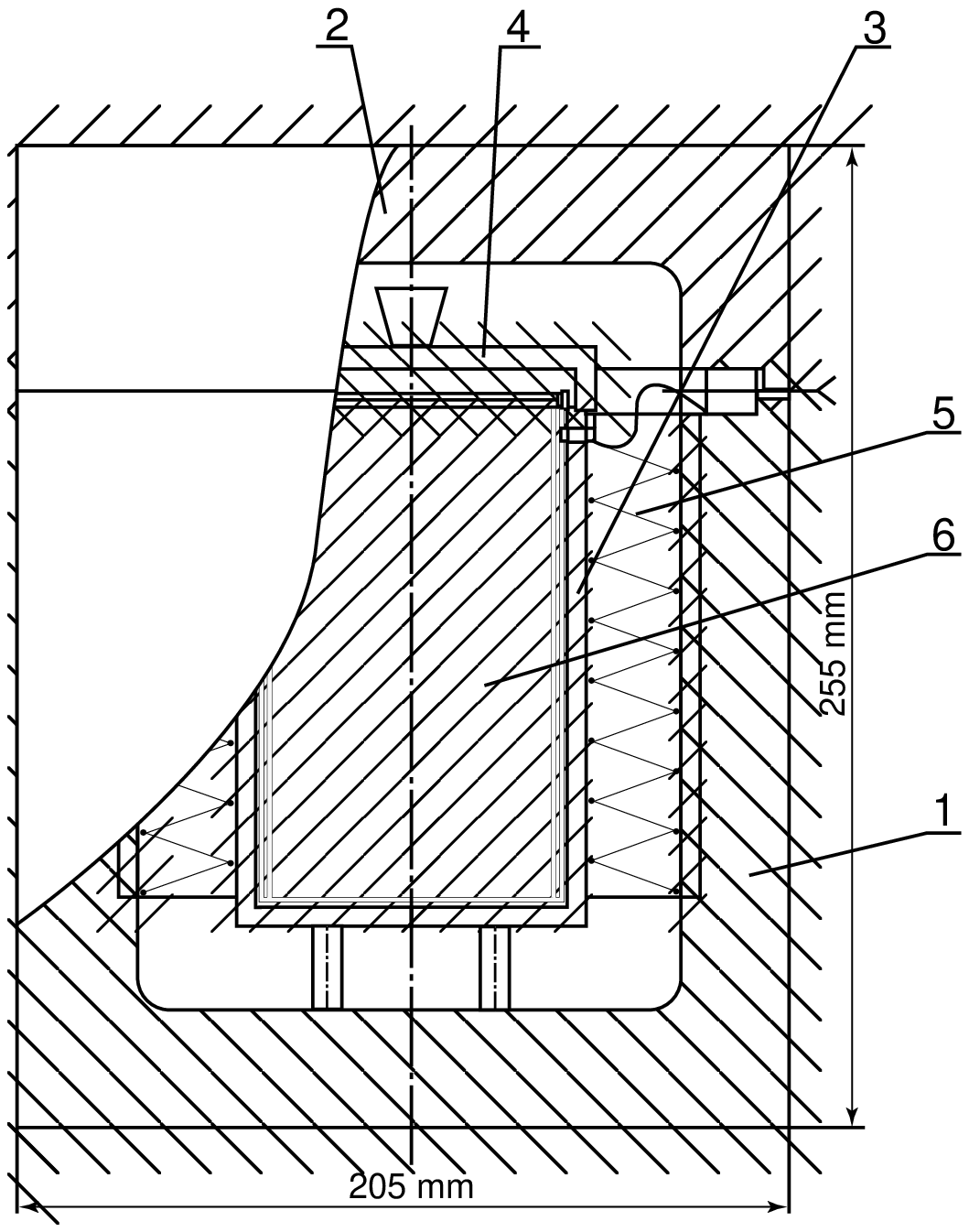}
\caption{Schematic drawing of the calorimeter.  Individual parts
are (1) copper block, (2) lid of copper block, (3) copper cup,
(4) lid of copper cup, (5) thermopile, and (6) source or
electroheater.}
\label{calorimeter}
\end{center}
\end{figure}

     The special calorimeter shown in Fig.\ \ref{calorimeter} was
built \cite{Belousev et al. 91} to measure the heating power from
the \nuc{51}{Cr} source.  It consisted of two identical
calorimetric transducers located side by side.  The internal
section of each transducer, into which the \nuc{51}{Cr} source
was placed, was a copper cup 95 mm in diameter and 150 mm high,
with a wall thickness of 5 mm.  The copper cup was inside the air
cavity of a large 68-kg copper block.  A thermopile consisting of
120 Chromel-Alumel thermocouples connected in series was placed
between the cup and the internal wall of the copper block.
Thermocouple hot junctions were distributed evenly over the cup
surface; the cold junctions were fixed to the internal surface of
the copper block.  The voltage produced by the thermopile was
thus proportional to the temperature difference between the cup
and its copper block.  The heat produced by the source was quite
large; to improve the heat exchange eight copper plates were
placed between the cup and the copper block.  The power of the
source warmed the cup and provided heat that was transferred to
the copper block.  This can be expressed as
     \begin{equation}
     \label{calorimeter operation}
     P = c \frac{dT_s}{dt} + K \Delta T,
     \end{equation}
\noindent where $P$ is the heat power of the source (W), $c$ is
the heat capacity of the source and the cup (J/\degreesC), $T_s$
is the temperature of the source and cup (\degreesC), $t$ is the
time (s), $K$ is the heat-transfer coefficient [J/(\degreesC s)],
and $\Delta T$ is the temperature difference between the cup and
the copper block (\degreesC).  The first term in
Eq.~(\ref{calorimeter operation}) represents the heating of the
source-cup system; the second term describes the transfer of heat
to the copper block.

     To understand the operation of the calorimeter, consider a
typical measurement.  When the source was first put into the cup,
$\Delta T = 0$; so all heat from the source served only to warm
the cup.  Then, as the temperature of the cup increased, heat
began to be transferred to the copper block and the heating rate
of the cup containing the source decreased.  When thermal
equilibrium was reached, which required approximately 6 h, the
cup-source system was at a constant temperature and all heat
produced by the source was transferred to the copper block.  In
this condition the signal from the thermopile was constant, and
the source thermal power, by Eq.~(\ref{calorimeter operation}),
was determined only by the temperature difference $\Delta T$ and
the heat-transfer coefficient $K$.

     The-heat transfer coefficient $K$ was determined using
electroheaters made from steel or aluminum whose outside
dimensions coincided exactly with the outside dimensions of the
source.  The heater power was varied by controlling the current
to an internal Nichrome or constantan winding.  Each heater was
used for calibration up to its maximum power.

\begin{figure}[ht]
\begin{center}
\includegraphics*[width=3.375in,bb=21 9 437 286] {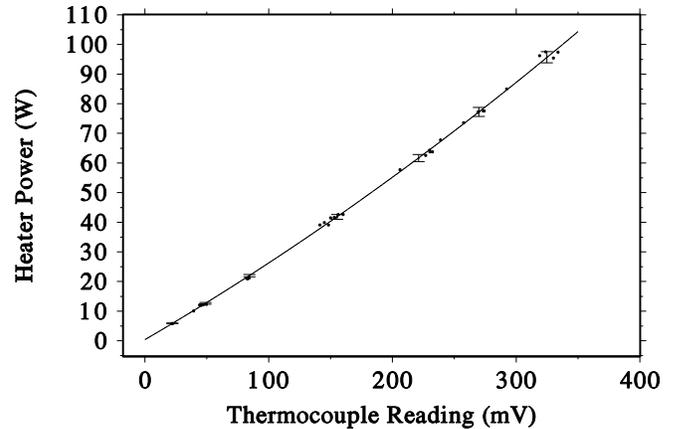}
\caption{Calibration curve of the calorimeter.  The thermocouple
readings and the inferred power for the seven measurements of the
Cr source are also indicated.}
\label{calorimeter calibration curve}
\end{center}
\end{figure}

     The calibration curve of the thermistor reading as a
function of heater power in watts is shown in
Fig.\ \ref{calorimeter calibration curve}.  The uncertainty
associated with each measurement was approximately \+/-2\%.  A
fit to the calibration curve with a second-order polynomial gave
the result
     \begin{equation}
     P = 0.43(14) + 0.2418(41) V + 0.000\:159(16) V^2,
     \end{equation}
\noindent where $P$ is the power in watts and $V$ is the
thermistor reading in mV.  The numbers in parentheses represent
the uncertainties in the final digits of each parameter.  With
each measurement weighted by the 2\% uncertainty, $\chi^2$ for
the fit was 36.9 for 35 data values.

\begin{table}[ht]
\caption{Source power measurements with the calorimeter.}
\label{source activity vs time}
\begin{tabular}{d d c c}
Days after 18:00                          &  Thermocouple  &         
Deduced            & Power on 26  \\
\multicolumn{1}{c}{on 26 Dec.\ 1994}      &  voltage (mV)  &         
power (W)          &Dec.\ 1994 (W)\\
\hline
6.62               &    324.8     & 95.7 \+/- 1.9 &    112.9 \+/-
2.2                \\
13.90              &    269.9     & 77.2 \+/- 1.5 &    109.3 \+/-
2.2                \\
23.08              &    221.2     & 61.6 \+/- 1.2 &    109.8 \+/-
2.2                \\
39.08              &    155.36    & 41.8 \+/- 0.8 &    111.1 \+/-
2.2                \\
66.21              &    84.46     & 22.0 \+/- 0.4 &    115.1 \+/-
2.3                \\
88.17              &    48.88     & 12.6 \+/- 0.3 &    114.5 \+/-
2.3                \\
118.17             &    22.20     &  5.9 \+/- 0.1 &  113.0 \+/- 2.3
\end{tabular}
\end{table}

\begin{figure}[ht]
\begin{center}
\includegraphics*[width=2.89in,bb=6 -2 284 353] {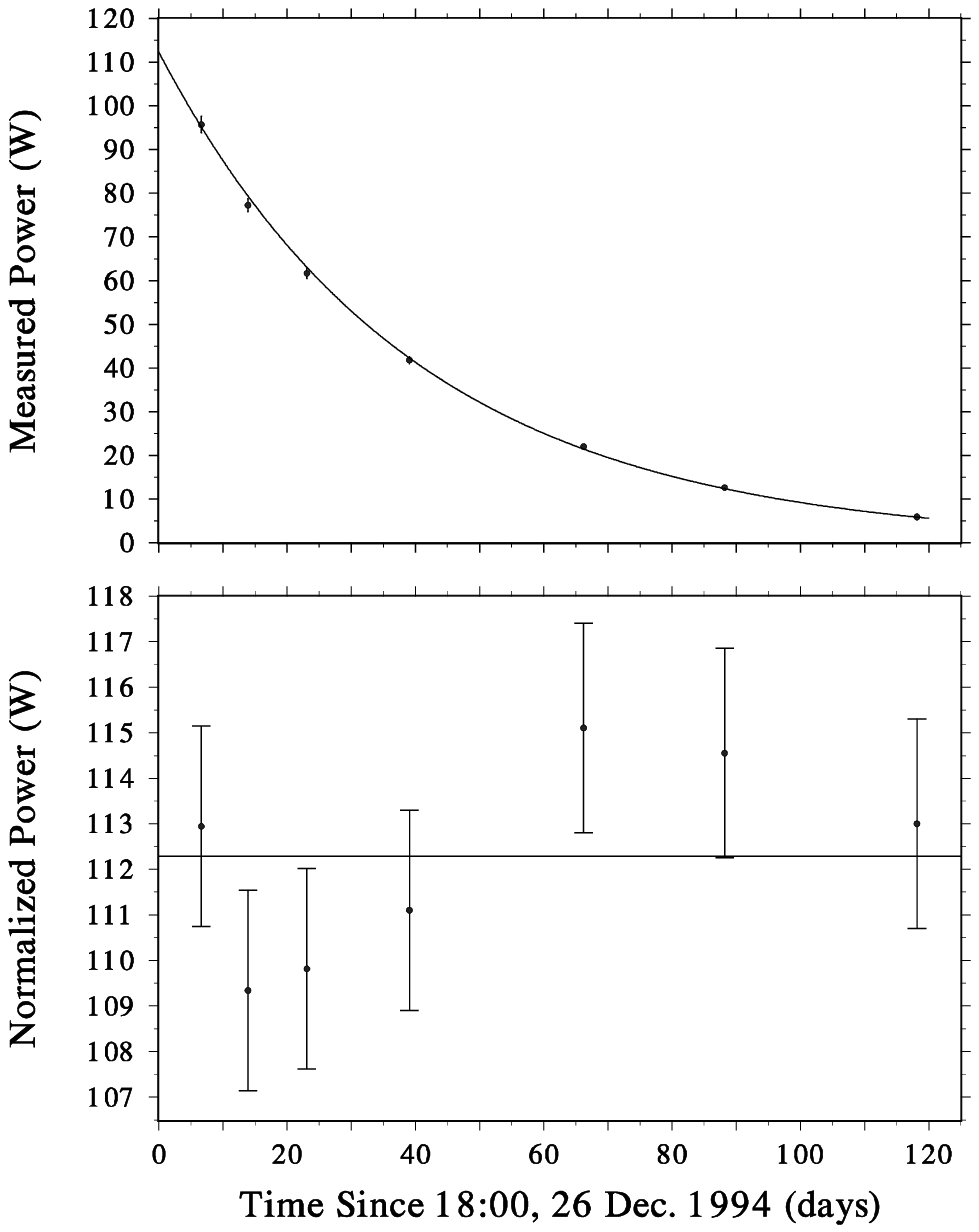}
\caption{The seven individual source activity measurements.  The
line is a weighted fit to the data points with an exponential
function whose half-life is that of \nuc{51}{Cr}.  In the lower
panel the power is normalized to 18:00 on 26 December 1994.}
\label{activity vs time}
\end{center}
\end{figure}

     The heat produced by the \nuc{51}{Cr} source was measured
between extractions for a total of seven measurements.  The
results of these measurements are shown numerically in Table
\ref{source activity vs time} and graphically in
Fig.\ \ref{activity vs time}.  The uncertainty in each
measurement is only that propagated from the calibration curve.
Each value is normalized to the activity on 26 Dec.\ at 18:00
taking into account the decay of the \nuc{51}{Cr}.  A weighted
average of these seven power measurements gives a value on 26
Dec.\ at 18:00 of 112.3~\+/- 0.8 W
(Fig.\ \ref{activity vs time}).  $\chi^2$ for this average is
6.0.  As a test, we performed this same fit, allowing the
parameter associated with the \nuc{51}{Cr} half-life to vary.
The best fit half-life was determined to be 28.03~\+/- 0.23 days,
in reasonable agreement with the known value of 27.702~\+/- 0.004
days \cite{Table of Isotopes 96,Zhou 91}.

     The decay of \nuc{51}{Cr} gives an average energy release of
36.67~\+/- 0.20 keV/decay.  Using 1.6022 \E{-19} (W s)/eV and 3.7
\E{10} decays/(Ci s), this implies a conversion factor of
4.600~\+/- 0.025 kCi \nuc{51}{Cr}/W.  The \nuc{51}{Cr} activity
at the time it was first placed in the reactor containing Ga was
thus 516.6~\+/- 3.7 kCi, where the uncertainty is entirely
statistical.

     Several systematic uncertainties are associated with this
source activity determination.  Before taking a thermocouple
measurement we waited for 12 h to be sure that the source (or
electroheater) and the copper block in the calorimeter were in
thermal equilibrium.  It is estimated that the uncertainty due to
different stabilization times between the source and the
calibration heaters can be no more then 0.6\% or 3.1 kCi.

     The 0.54\% uncertainty in the energy released per
\nuc{51}{Cr} decay leads directly to an uncertainty of \+/-2.8
kCi in the source activity.  We should note that the value we use
for the energy release differs slightly from the value of
36.510~\+/- 0.161 keV/decay used in Ref.~\cite{Anselmann et al.
95}.  There are two primary differences between these
calculations: First, Ref.~\cite{Anselmann et al. 95} used a
branching ratio to the 320-keV level of 0.0986 \cite{Schotzig and
Schrader 93}, whereas we chose to use 0.0988 \cite{Table of
Isotopes 96}.  Second, Ref.~\cite{Anselmann et al. 95} ignored
the contribution of internal bremsstralung, which contributes
approximately 96 eV/decay to the average \cite{Table of
Radioactive Isotopes 86}, whereas we have included it here.

     The half-life of Cr is known to 0.02\%.  To estimate how
large an uncertainty this introduced in our source activity
estimate, we repeated the fit to the Cr decay curve using a value
for the half-life which differed from the known value by one
standard deviation.  This changed the power determination by
0.2\% or 1.0 kCi and we take that as an estimate of the related
uncertainty.

     Radioactive impurities in the source can also give rise to
heat which would be incorrectly attributed to \nuc{51}{Cr}.  The
impurity content of the source was considered above in Section
\ref{source impurities}, and the contribution of each impurity to
the source power is given in Table \ref{nuclide impurities}.  The
effective Cr activity from all impurities was only 100 Ci at the
reference time of 18:00 on 26 Dec.\ 1994, which is a completely
negligible 0.02\% uncertainty.  Because the half-life of the
impurities was longer than that of Cr, the fractional size of
this error increased with time.  For the final calorimeter
measurement on 24 April 1995, the fraction had risen to 0.14\%.

\begin{table*}[ht]
\caption{Summary of the uncertainties associated with the source
activity as deduced from the calorimetry data.  In the total, the
uncertainty due to contamination is taken to be the larger of the
two extremes.  All uncertainties are symmetric.}
\label{source activity uncertainties}
\begin{tabular}{l d d}
                                                        &
\multicolumn{2}{c}{Uncertainty}                   \\ \cline{2-3}
Origin of uncertainty                                   &
Percentage                                              &
Magnitude (kCi)                                         \\
\hline
Statistics (112.3 \+/- 0.8 watts)                       &       0.8
&                                                      3.9     \\
Calorimeter equilibration                               &       0.6
&                                                      3.1     \\
Power to activity conversion (4.600 \+/- 0.025 kCi/W)   &       0.54
&                                                      2.8     \\
\nuc{51}{Cr} half-life (27.702 \+/- 0.004 days)         &       0.2
&                                                      1.0     \\
Contamination (26 Dec.\ 1994)                           &       0.02
&                                                      0.10    \\
Contamination (24 April 1994)                           &       0.14
&                                                      0.72    \\
\\
Total uncertainty (added in quadrature)                 &       1.2
&                                                      6.0
\end{tabular}
\end{table*}

     Other possible contributions to the systematic uncertainty
in the calorimetric determination of the source activity have
been considered, such as the escape of some of the 320-keV gamma
rays of \nuc{51}{Cr} from the source.  All such contributions
were shown to be negligible.  Table
\ref{source activity uncertainties} summarizes the various
components of the source activity uncertainty that were described
above.  Adding the statistical and systematic components in
quadrature gives the final value of 516.6~\+/- 6.0 kCi at the
reference time.

     The following subsections describe other independent methods
used to measure the source activity.  The calorimeter technique
is the most precise and we use its result; the other methods add
confidence in the calorimetric determination.

\subsection{Source activity from direct counting}

     This section describes an independent determination of the
source activity that used a Ge(Li) detector to measure the
320-keV gamma rays emitted by \nuc{51}{Cr}.  Because of the high
initial activity of the source, these measurements could only be
carried out after the gallium exposures at the Baksan Neutrino
Observatory had finished and the source had been returned to
Aktau.  At that time the \nuc{51}{Cr} activity had decreased by a
factor of more than 1000.

     The procedure for these measurements consisted of two steps:
first, the relative activity of all 44 Cr rods was measured, and
second, the absolute activity of a single monitor rod was
determined.  For the first step, two collimators were installed
in the hot chamber of BN-350.  A chromium rod was placed in a
special transit in front of the slit of the first collimator.
The transit moved in a vertical direction using a manipulator of
the hot chamber and contained a motor which rotated the rod
during measurement.  The position of the Cr rod in relation to
the collimator slit was controlled by electronic readout of the
manipulator and by visual observation.  Gamma rays passed through
the slit of the second collimator and were counted by a Ge(Li)
detector outside the hot cell.  The activity of each rod was
measured at three points along its length and the angular
distribution was averaged because of the rotation of the rod.
This system provided the average value of the activity of all Cr
rods.  The uncertainty in the relative activity of one rod was
1\% and was determined by statistics, background, and the
stability of the measurement geometry.  The uncertainty in the
sum of the relative activities of all rods was added in
quadrature, resulting in an uncertainty of 0.3\%.

     The second step was to measure the absolute activity of the
monitor rod.  This rod was completely dissolved in HCl acid.  A
small portion of this solution was diluted to prepare samples and
their activity was measured.  The uncertainty in activity from
differences in sampling procedure was 3\%; there was also a 1\%
error in volume because of the successive dilutions.

     The standard deviation of a set of measurements of the count
rate in the \nuc{51}{Cr} photopeak had an uncertainty of 1.2\%
due to statistics, background, and a dead time correction factor.
The uncertainty in the efficiency of the detector was 3\%.  The
ratio of the mass of the monitor rod to the mass of all rods in
the source had a 1\% uncertainty.  The quadratic sum of all these
uncertainties was 4.7\%.  The final result of this method of
source activity measurement is 510~\+/- 24 kCi at our reference
time (18:00 on 26 Dec.\ 1994).

\subsection{Source activity from reactor physics}

     The source activity can be determined, in principle, by
direct neutron transport calculations using the geometry of the
reactor and the irradiation assemblies.  Such a calculation has
many difficulties which limit its precision.  Based on test
results from the experimental reactor at Obninsk and irradiation
of a small mass of \nuc{50}{Cr} in the reactor at Aktau, the
calculated activity was 554~\+/- 55 kCi at our reference time, in
agreement with the calorimeter measurement.

\section{Counting of $^{\bf71}$G\lowercase{e}}
\label{counting}

\begin{table*}
\caption{Counting parameters.  $\Delta$ is the exponentially
weighted live time after all time cuts have been applied.  The
second extractions were not counted in an electronics system with
a digitizer so $L$-peak analysis could not be performed.  There
are no entries for Cr 2-2 as the counter failed and for Cr 7-2 as
the sample was not counted.}
\label{counting parameters}
\begin{tabular}{l d d d d d d c c c c}
           &\multicolumn{2}{c}{Counter filling}     &
\multicolumn{2}{c}{Counter efficiency before}       &
\multicolumn{2}{c}{Day counting}                    &
\multicolumn{2}{c}{Live time of}                    &            
     \\ \cline{2-3}
Extraction & GeH$_4$         &Pressure              &
\multicolumn{2}{c}{rise time or energy cuts}        &
\multicolumn{2}{c}{began in 1995}                   &
\multicolumn{2}{c}{counting (days)}   &\multicolumn{2}{c}{$\Delta$}
\\\cline{4-5} \cline{6-7} \cline{8-9} \cline{10-11}
name       &fraction (\%)    &(mm Hg)               &  $L$ peak &
$K$ peak   &$L$ peak         &$K$ peak              &  $L$ peak &
$K$ peak   &$L$ peak         &$K$ peak              \\
\hline
Cr 1       &      6.5 &    690 &                  0.335   &   0.356
&         5.82    &          2.90                   &   137.8   &
142.1      &      0.618      &0.734                 \\
Cr 2       &      8.0 &    685 &                  0.326   &   0.344
&        10.42    &         10.35                   &   134.5   &
136.7      &      0.753      &0.767                 \\
Cr 3       &      7.5 &    650 &                  0.329   &   0.338
&        19.41    &         19.34                   &   104.0   &
105.6      &      0.792      &0.804                 \\
Cr 4       &      8.5 &    665 &                  0.329   &   0.341
&        35.39    &         35.32                   &   132.3   &
134.8      &      0.824      &0.837                 \\
Cr 4-2     &     13.5 &    650 &                   ---    &   0.321
&         ---     &         37.24                   &    ---    &
126.6      &    ---          &0.584                 \\
Cr 5       &      7.5 &    650 &                  0.350   &   0.359
&        61.52    &         61.52                   &   119.6   &
122.0      &      0.760      &0.774                 \\
Cr 6       &      8.7 &    645 &                  0.359   &   0.365
&        84.44    &         84.38                   &   120.7   &
123.4      &      0.506      &0.521                 \\
Cr 6-2     &     13.5 &    695 &                   ---    &   0.350
&         ---     &         86.21                   &    ---    &
152.0      &    ---          &0.841                 \\
Cr 7       &      7.0 &    645 &                  0.329   &   0.337
&        114.49   &        114.29                   &   129.8   &
132.4      &      0.775      &0.784                 \\
Cr 8       &      7.0 &    700 &                  0.324   &   0.347
&        145.41   &        145.41                   &   151.8   &
154.9      &      0.729      &0.747
\end{tabular}
\end{table*}

     The number of \nuc{71}{Ge} atoms extracted from the gallium
was determined by the same procedure as used for solar neutrino
measurements.  Very briefly, the extracted Ge was synthesized
into the counting gas GeH$_4$, mixed with Xe, and inserted into a
very-low-background proportional counter.  All pulses from this
counter were then recorded for about the next 6 months.  The
counter body was made from synthetic quartz and cathode from
ultrapure Fe; the volume was about 0.75 cm$^3$.  To detect and
suppress background, the counter was placed in the well of a
large NaI detector, which was in turn contained within a massive
Cu, W, Pb, and Fe passive shield.  The parameters of counting are
given in Table \ref{counting parameters}.

     The data recording system made hardware measurements of the
pulse energy, ADP (a parameter inversely proportional to the
pulse rise time during the first few ns), energy and time of any
NaI events that occurred within $-8$ ms to +8 ms of the counter
pulse, and event time.  In addition, all the first extractions
were measured in a counting system that digitized the pulse
waveform for 800 ns after pulse onset.

     \nuc{71}{Ge} decays by electron capture with an 11.4-day
half life and emits Auger electrons and x rays whose sum energy
is usually either 10.4 keV (the $K$ peak) or 1.2 keV (the $L$
peak).  The radial extent of these low-energy electrons in the
counter is very short, producing a pulse waveform with a fast
rise time.  Background events, such as a minimum ionizing
particle that traverses the counter, may deposit a similar amount
of energy in the counter gas, but will usually have longer radial
extent and hence slower rise time.  Measurement of the rise time
thus gives a very powerful suppression of background.  For all
first extractions the rise time was determined by fitting the
digitized waveform to an analytical formula \cite{Elliott 90}
that describes the pulse shape in terms of the radial extent of
the trajectory in the counter.

     The counters had a hole in the cathode near the center of
the active volume with a thin section in the quartz envelope so
the gas filling could be directly irradiated with the 5.9-keV x
rays from \nuc{55}{Fe}.  Counters were calibrated with
\nuc{55}{Fe} just before the start of counting, about 3 days
later, 1 week later, and then at 2--3 week intervals until
counting ended.  At least four \nuc{55}{Fe} calibrations were
made for each run during the first month of counting, while the
\nuc{71}{Ge} was decaying.  For all eight first extractions the
average change in the \nuc{55}{Fe} peak position during this time
was 2.4\%.  They were also calibrated with \nuc{109}{Cd} which
fluoresced the Fe cathode and made 6.4-keV x rays throughout the
counter volume.  For these \nuc{109}{Cd} calibrations, the source
was positioned so that it did not see the side hole in the
cathode; the peak position was thus representative of the
response of the entire counter.  By comparing the predicted
position of the 6.4-keV peak based on the \nuc{55}{Fe}
calibration with the actual position in the \nuc{109}{Cd}
calibration, a correction factor was derived that modified the
energy scale from the \nuc{55}{Fe} calibration to account for any
polymerization that might be present on the anode wire in the
vicinity of the side hole.  For the eight first extractions this
correction averaged 4.5\% with a range from 0\% to 11\%.

     After the counting of the samples from the Cr experiment was
completed, in the fall of 1995, measurements of the counting
efficiency were made.  Two different techniques and three
different isotopes were employed: \nuc{37}{Ar} to measure volume
efficiency, and \nuc{69}{Ge} and \nuc{71}{Ge} to measure the $L$-
and $K$ peak efficiencies \cite{Abdurashitov et al. 95}.  The
volume efficiency of all counters used for first extractions was
directly measured with \nuc{37}{Ar}.  The calculated counting
efficiency, using the measured pressure, GeH$_4$ fraction, and
\nuc{37}{Ar} volume efficiency, is given for each extraction in
Table \ref{counting parameters}.  The total uncertainty in these
calculated efficiencies is 3.1\%.

\section{Data Analysis and Results}

\subsection{Event selection}
\label{event selection}

\begin{figure}[t]
\begin{center}
\includegraphics*[width=3.375in,bb=0 -4 380 358] {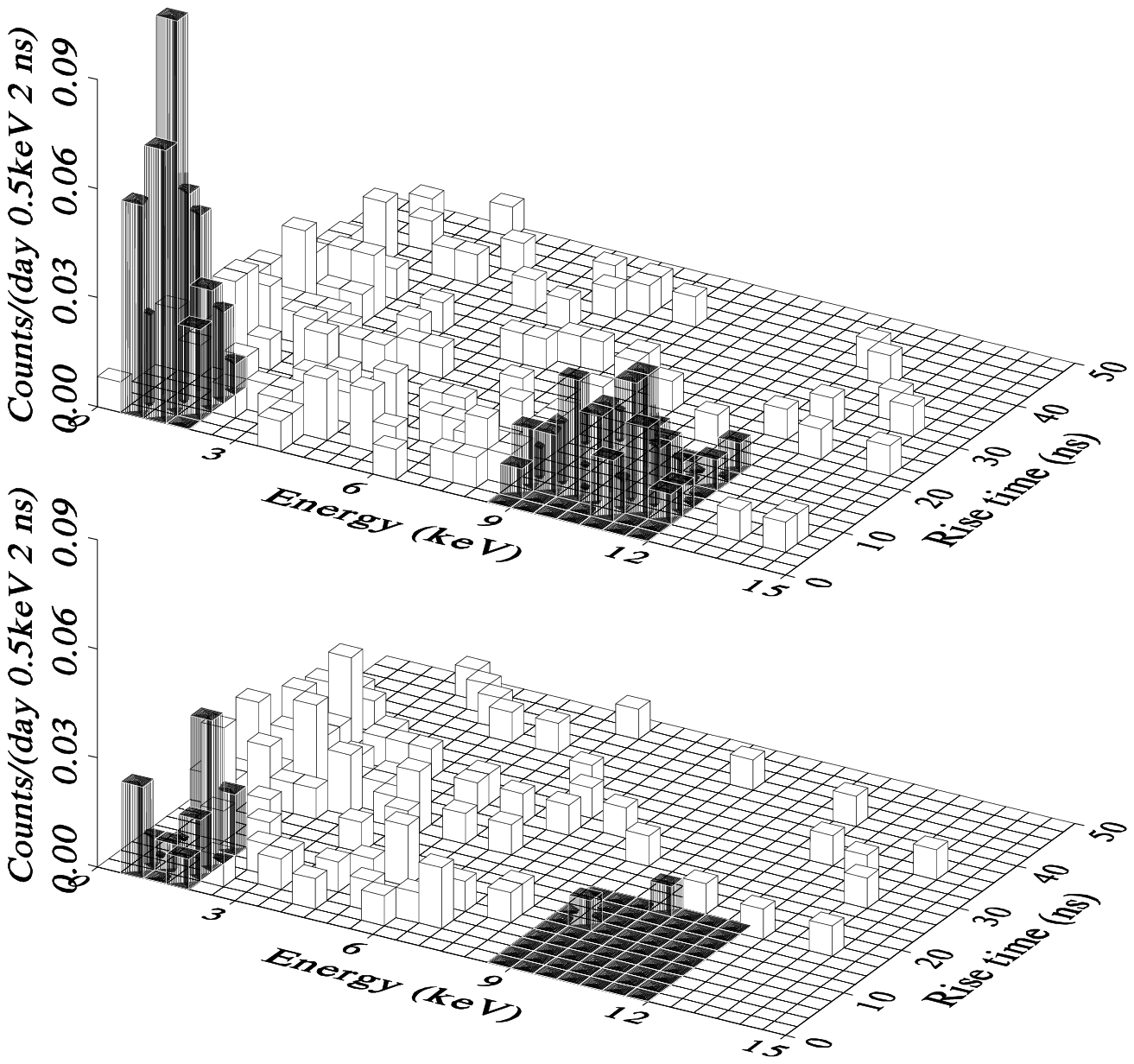}
\caption{Upper panel shows the energy rise time histogram of all
events observed during the first 30 days after extraction for the
first five Cr exposure measurements.  The live time is 120.1
days.  The expected location of the \nuc{71}{Ge} $L$ and $K$
peaks is shown darkened.  Lower panel shows the same histogram
for all events that occurred during an equal live time interval
at the end of counting.}
\label{2D histogram}
\end{center}
\end{figure}

     Candidate \nuc{71}{Ge} events were selected in exactly the
same manner as in our extractions to measure the solar neutrino
capture rate.  The first step was to apply time cuts to the data
that serve to suppress false \nuc{71}{Ge} events that may be
produced by Rn outside the proportional counter and by Rn added
to the counter during filling.  In the next step, events that
were in coincidence with the surrounding NaI counter were
eliminated.  For the first five extractions, a histogram of the
250 events that remained which occurred during the first 30 days
after extraction is given in the upper panel of Fig.\ \ref{2D
histogram}.  The darkened areas are the locations of the
\nuc{71}{Ge} $L$ and $K$ peaks as predicted from the \nuc{55}{Fe}
and \nuc{109}{Cd} calibrations.  For comparison, an identical
spectrum of the 113 events that occurred in these extractions
during an interval of equal live time at the end of counting
(more than 122 days after extraction) is given in the lower panel
of Fig.\ \ref{2D histogram}.  The \nuc{71}{Ge} $L$ and $K$ peaks
are very obvious in the spectrum at the beginning of counting,
but are absent in the spectrum at the end of counting because the
\nuc{71}{Ge} has decayed away.  The number of counts outside the
two peaks is approximately the same in both spectra because they
were produced by background processes.

     Windows with 98\% acceptance in energy [2 full width at half
maximum (FWHM) width] and 95\% acceptance in rise time (0--10 ns
in the $L$ peak and 0--18.4 ns in the $K$ peak) were then set
around the $L$ and $K$ peaks.  All events inside these windows
during the entire period of counting were considered as candidate
\nuc{71}{Ge} events.

\subsection{Maximum likelihood analysis}
\label{maximum likelihood analysis}

     The time sequence of the candidate \nuc{71}{Ge} events was
analyzed with a maximum likelihood method \cite{Cleveland 83} to
separate the \nuc{71}{Ge} 11.4-day decay from a constant rate
background.  The only differences between this analysis and that
done for the solar neutrino runs are that one must account for
the decay of the \nuc{51}{Cr} during the period of exposure,
include a fixed term for solar neutrino background, and add a
carryover term arising from the \nuc{71}{Ge} that was not removed
because of the approximately 15\% inefficiency of the preceding
chemical extraction.

     The likelihood function ($\cal L$) for each extraction is
given by Eq.~(17) of Ref.~\cite{Cleveland 83},
     \begin{equation}
     {\cal L} = e^{-(bT_L + a\Delta /\lambda_{71})}
     \prod_{i=1}^N \left[ b + ae^{-\lambda_{71}t_i} \right],
     \end{equation}
\noindent where $b$ is the background rate, $T_L$ is the live
time of counting, $\lambda_{71}$ is the \nuc{71}{Ge} decay
constant, $\Delta$ is the probability that a \nuc{71}{Ge} atom
that is extracted will decay during a time that it might be
counted, and $t_i$ are the times of occurrence of the $N$
candidate events.

     The parameter $a$ contains contributions from the three
separate processes (Cr source neutrinos, solar neutrinos, and
carryover) that are able to give \nuc{71}{Ge} in each extraction,
i.e., $a = a_{\text{Cr}} + a_\odot + a_{\text{carryover}}$.  It
follows from Eq.~(11) and Eq.~(12) of Ref.~\cite{Cleveland 83}
that these three terms are given for extraction $k$ by

\begin{eqnarray}
\label{aCr}
a_{\text{Cr}}^k & = & p_{\text{Cr}} \epsilon^k
\,\exp[-\lambda_{51}(t_s^k -T)] \\
& & \times
[\exp(-\lambda_{51} \theta_{\text{Cr}}^k) -
\exp(-\lambda_{71} \theta_{\text{Cr}}^k)]
/(1 - \lambda_{51}\lambda_{71}), \nonumber
\end{eqnarray}

\begin{equation}
\label{asolar}
a_\odot^k = p_\odot \epsilon^k \,[1- \exp(-\lambda_{71}
\theta_\odot^k)],
\end{equation}

\begin{equation}
a_{\text{carryover}}^k = a^{k-1}
\frac{\epsilon^k}{\epsilon^{k-1}}
\,\exp(-\lambda_{71} \theta_\odot^k)[1 -
\epsilon_{\text{Ga}}^{k-1}].
\end{equation}

\noindent Here $p_{\text{Cr}}$ and $p_\odot$ are the rates of
production of \nuc{71}{Ge} by the \nuc{51}{Cr} source and solar
neutrinos, respectively; $\lambda_{51}$ is the decay constant of
\nuc{51}{Cr}; $t_s$ is the starting time of each source exposure;
$T$ is the source activity reference time of 18:00 on 26 Dec.\
1994; $\theta_{\text{Cr}}$ and $\theta_\odot$ are the times of
exposure of the Ga to the \nuc{51}{Cr} source and to solar
neutrinos, respectively; $\epsilon$ is the product of extraction
and counting efficiencies; and $(1-\epsilon_{\text{Ga}})$ is the
inefficiency of extraction of Ge from the Ga.  With these
definitions, as the source decays, the production rate
$p_{\text{Cr}}$ is automatically referred to time $T$.

     In the maximization procedure to obtain $p_{\text{Cr}}$ for
each run, the solar production rate $p_\odot$ was held fixed at
0.27/day, the rate corresponding to 69 SNU \cite{Abdurashitov et
al. 94} on 13.1 tonnes of Ga.  Since second extractions followed
extractions 2, 4, 6, and 7, the carryover correction was only
applied to extractions 2, 4, and 6.  Errors on $p_{\text{Cr}}$
with one sigma confidence were set by finding the values of the
rate that decreased the likelihood function from its value at the
maximum by the factor $e^{-0.5}$.  For each test value of
$p_{\text{Cr}}$ during this search, all the variables in the
likelihood function except $p_{\text{Cr}}$ were maximized.  The
overall production rate from the Cr source
$p_{\text{Cr}}^{\text{global}}$ was obtained by maximizing the
product of the likelihood functions for each run.  In these
maximizations the background rates in the $L$ and $K$ peaks for
each run were free parameters.

\subsection{Results}
\label{results}

\begin{figure}[t]
\begin{center}
\includegraphics*[width=2.73in,bb=9 -1 271 356] {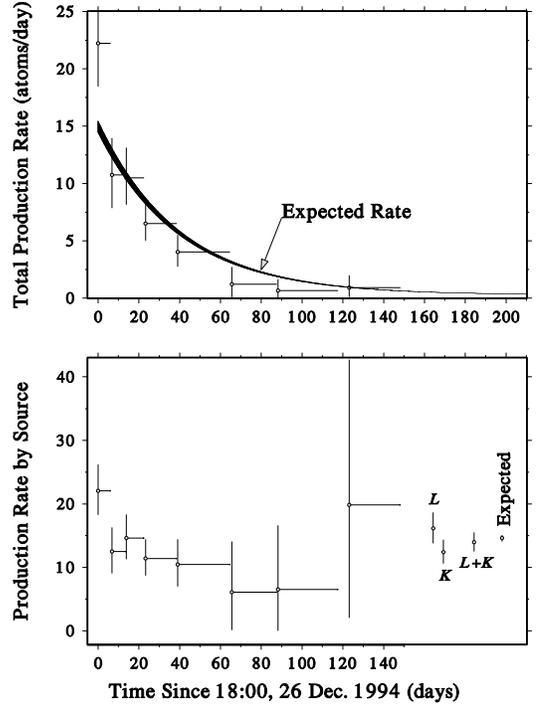}
\caption{The eight \nuc{71}{Ge} production rate measurements.
The horizontal lines indicate the beginning and ending of each
exposure with the vertical lines showing the measured production
rate and its statistical error.  The upper panel shows the total
\nuc{71}{Ge} production rate from the source and from solar
neutrinos.  The expected rate calculated from the 517 kCi source
activity and the cross section of Bahcall \protect\cite{Bahcall
97} is shown darkened.  The lower panel shows only the production
rate from the \nuc{51}{Cr} source, where each rate has been
normalized to the time of the start of the first exposure.  The
combined results of all measurements are shown at the right, with
the $L$-peak, $K$ peak, and $L$- plus $K$ peak results shown
separately.  The expected production rate and its uncertainty are
shown at the extreme right.}
\label{rate vs time}
\end{center}
\end{figure}

\begin{table*}[ht]
\caption{Results of analysis of $L$-peak events selected by pulse
shape.  The production rate for the individual exposures is
referred to the starting time of each exposure.  The production
rate for the combined result is referred to the time of the start
of the first exposure.  The second extractions were not counted
in an electronics system with a digitizer so event selection
based on pulse shape could not be made.  The parameter $Nw^2$
measures the goodness of fit of the sequence of event times
\protect\cite{Marshall 58,Cleveland 98}.  The probability was
inferred from $Nw^2$ by simulation.}
\label{L results}
\begin{tabular}{l d d d d d c c d}
           & Number of &    Number    &\multicolumn{3}{c}{Number
of events assigned to} &\nuc{71}{Ge} production rate        &       
&           \\ \cline{4-6}
Extraction & candidate &    fit to    & \nuc{51}{Cr} source &
Solar $\nu$&           &by \nuc{51}{Cr} source   &                 &
Probability\\
name       &   events  & \nuc{71}{Ge} &      production     &
production & Carryover & (atoms/day)  &        $Nw^2$       &
(percent)  \\
\hline
Cr 1       &    23.    &     20.9     &         22.5        &     0.4
&          0.    &28.5 $^{+6.6}_{-6.8}$          &        0.173    &
24.        \\
Cr 2       &    22.    &     11.9     &         10.4        &     0.3
&         1.1    &10.8 $^{+6.0}_{-2.9}$          &        0.036    &
81.        \\
Cr 3       &    22.    &     11.9     &         11.4        &     0.5
&          0.    &10.0 $^{+4.4}_{-3.5}$          &        0.062    &
43.        \\
Cr 4       &    24.    &     15.1     &         13.8        &     0.6
&         0.7    &8.0 $^{+3.8}_{-1.8}$&        0.082        &     37.
\\
Cr 5       &    20.    &     8.8      &         7.9         &     0.9
&          0.    &4.3 $^{+2.8}_{-1.7}$&        0.079        &     31.
\\
Cr 6       &    34.    &     0.6      &         0.0         &     0.5
&         0.2    &0.0 $^{+3.3}_{-0.0}$&        0.045        &     82.
\\
Cr 7       &    14.    &     2.9      &         2.1         &     0.8
&          0.    &1.2 $^{+2.0}_{-0.7}$&        0.118        &     23.
\\
Cr 8       &    11.    &     2.8      &         2.2         &     0.7
&          0.    &1.4 $^{+2.0}_{-1.1}$&        0.067        &     50.
\\
\\
Combined   &    170.   &     78.2     &         71.6        &     4.7
&         1.9    &16.1 $^{+2.5}_{-2.3}$          &        0.104    &
25.
\end{tabular}
\end{table*}

\begin{table*}[ht]
\caption{Results of analysis of $K$ peak events selected by pulse
shape.  See caption for Table \protect\ref{L results} for further
explanation.}
\label{K results}
\begin{tabular}{l d d d d d c c d}
           & Number of &    Number    &\multicolumn{3}{c}{Number
of events assigned to} &\nuc{71}{Ge} production rate        &       
&           \\ \cline{4-6}
Extraction & candidate &    fit to    & \nuc{51}{Cr} source &
Solar $\nu$&           &by \nuc{51}{Cr} source   &                 &
Probability\\
name       &   events  & \nuc{71}{Ge} &      production     &
production & Carryover & (atoms/day)  &        $Nw^2$       &
(percent)  \\
\hline
Cr 1       &    20.    &     16.4     &         15.9        &     0.5
&          0.    &17.2 $^{+5.1}_{-4.7}$          &        0.035    &
90.        \\
Cr 2       &    18.    &     12.2     &         10.6        &     0.4
&         1.2    &10.2 $^{+5.4}_{-2.9}$          &        0.319    &
3.         \\
Cr 3       &    18.    &     13.2     &         12.7        &     0.5
&          0.    &10.5 $^{+3.7}_{-2.8}$          &        0.515    &
1.         \\
Cr 4       &    12.    &     10.4     &         9.1         &     0.6
&         0.7    &5.0 $^{+2.5}_{-1.4}$&        0.060        &     69.
\\
Cr 5       &    15.    &     7.9      &         6.9         &     0.9
&          0.    &3.6 $^{+2.3}_{-1.2}$&        0.034        &     84.
\\
Cr 6       &     8.    &     2.8      &         2.1         &     0.5
&         0.2    &1.6 $^{+2.3}_{-1.0}$&        0.041        &     79.
\\
Cr 7       &    12.    &     1.0      &         0.1         &     0.9
&          0.    &0.1 $^{+1.7}_{-0.1}$&        0.071        &     60.
\\
Cr 8       &    10.    &     2.0      &         1.2         &     0.7
&          0.    &0.7 $^{+1.6}_{-0.5}$&        0.064        &     59.
\\
\\
Combined   &    113.   &     67.5     &         60.3        &     5.1
&         2.1    &12.4 $^{+2.0}_{-1.8}$          &        0.042    &
87.
\end{tabular}
\end{table*}

\begin{table*}[ht]
\caption{Results of combined analysis of $L$- and $K$ peak events
selected by pulse shape.  See caption for Table
\protect\ref{L results} for further explanation.}
\label{KL results}
\begin{tabular}{l d d d d d c c d}
           & Number of &    Number    &\multicolumn{3}{c}{Number
of events assigned to} &\nuc{71}{Ge} production rate        &       
&           \\ \cline{4-6}
Extraction & candidate &    fit to    & \nuc{51}{Cr} source &
Solar $\nu$&           &by \nuc{51}{Cr} source   &                 &
Probability\\
name       &   events  & \nuc{71}{Ge} &      production     &
production & Carryover & (atoms/day)  &        $Nw^2$       &
(percent)  \\
\hline
Cr1        &    43.    &     36.9     &         36.0        &     0.9
&          0.    &22.0 $^{+4.1}_{-3.8}$          &        0.121    &
35.        \\
Cr2        &    40.    &     24.0     &         21.1        &     0.7
&         2.3    &10.5 $^{+3.2}_{-2.9}$          &        0.202    &
3.         \\
Cr3        &    40.    &     25.2     &         24.2        &     1.0
&          0.    &10.3 $^{+2.6}_{-2.3}$          &        0.120    &
15.        \\
Cr4        &    36.    &     25.2     &         22.5        &     1.3
&         1.4    &6.4 $^{+1.7}_{-1.5}$&        0.061        &     61.
\\
Cr5        &    35.    &     16.4     &         14.6        &     1.8
&          0.    &3.9 $^{+1.5}_{-1.3}$&        0.034        &     84.
\\
Cr6        &    42.    &     4.1      &         2.8         &     0.9
&         0.3    &1.2 $^{+1.5}_{-1.0}$&        0.046        &     79.
\\
Cr7        &    26.    &     3.9      &         2.2         &     1.7
&          0.    &0.6 $^{+1.0}_{-0.6}$&        0.081        &     43.
\\
Cr8        &    21.    &     4.5      &         3.1         &     1.4
&          0.    &0.9 $^{+1.1}_{-0.8}$&        0.034        &     89.
\\
\\
Combined   &    283.   &    143.7     &        130.0        &     9.8
&         4.0    &14.0 $^{+1.5}_{-1.5}$          &        0.068    &
50.
\end{tabular}
\end{table*}

     The set of Tables \ref{L results}, \ref{K results}, and
\ref{KL results} gives the results of the data analysis for the
$L$ peak, $K$ peak, and $K+L$ peaks.  The result for the global
production rate in the combined fit to the eight extractions is
16.1~$^{+2.5}_{-2.3}$/day in the $L$ peak,
12.4~$^{+2.0}_{-1.8}$/day in the $K$ peak, and
14.0~$^{+1.5}_{-1.5}$/day in the $K+L$ peaks.  The uncertainties
here are all statistical.  Figure\ \ref{rate vs time} shows the
$K+L$ combined results for the eight exposures and extractions.
In the final three extractions only a few counts were produced by
the \nuc{51}{Cr} source; so these results for the global
production rates were almost unchanged if only the first five
extractions were used in the combined fit.

     A fit permitting the \nuc{71}{Ge} half-life to vary gave
13.5~\+/- 2.0 days, compared with its known half-life of 11.4
days.

\begin{table*}[ht]
\caption{Results of analysis of $K$ peak events selected by ADP.
The second extraction results were not used in the combined fit.}
\label{ADP results}
\begin{tabular}{l d d d d d c c d}
           & Number of &    Number    &\multicolumn{3}{c}{Number
of events assigned to} &\nuc{71}{Ge} production rate        &       
&           \\ \cline{4-6}
Extraction & candidate &    fit to    & \nuc{51}{Cr} source &
Solar $\nu$&           &by \nuc{51}{Cr} Source   &                 &
Probability\\
name       &   events  & \nuc{71}{Ge} &      production     &
production & Carryover & (atoms/day)  &        $Nw^2$       &
(percent)  \\
\hline
Cr 1       &    16.    &     16.0     &         15.5        &     0.5
&          0.    &15.3 $^{+4.0}_{-4.0}$          &        0.038    &
94.        \\
Cr 2       &    15.    &     10.8     &         9.3         &     0.4
&         1.2    &9.2 $^{+5.0}_{-2.6}$&        0.235        &      6.
\\
Cr 3       &    16.    &     12.9     &         12.4        &     0.5
&          0.    &10.4 $^{+3.7}_{-2.8}$          &        0.466    &
2.         \\
Cr 4       &     9.    &     9.0      &         7.7         &     0.6
&         0.7    &4.3 $^{+2.4}_{-1.0}$&        0.055        &     84.
\\
Cr 4-2     &     7.    &     0.2      &         0.1         &     0.1
&          0.    &0.1 $^{+2.0}_{-0.1}$&        0.219        &     12.
\\
Cr 5       &    13.    &     5.6      &         4.7         &     0.9
&          0.    &2.5 $^{+2.1}_{-1.0}$&        0.027        &     93.
\\
Cr 6       &     6.    &     1.8      &         1.2         &     0.5
&         0.2    &0.9 $^{+2.0}_{-0.8}$&        0.034        &     91.
\\
Cr 6-2     &     5.    &     0.1      &         0.0         &     0.1
&          0.    &0 $^{+1.0}_{-0.0}$  &        0.086        &     50.
\\
Cr 7       &     8.    &     1.9      &         1.0         &     0.9
&          0.    &0.6 $^{+1.6}_{-0.4}$&        0.038        &     85.
\\
Cr 8       &    11.    &     1.8      &         1.1         &     0.7
&          0.    &0.6 $^{+1.6}_{-0.5}$&        0.062        &     60.
\\
\\
Combined   &    94.    &     61.7     &         54.7        &     5.0
&         2.0    &11.2 $^{+1.8}_{-1.7}$          &        0.039    &
89.
\end{tabular}
\end{table*}

     Our solar neutrino results in the past have been based on
events selected by ADP.  Table \ref{ADP results} gives the
results of analysis of the Cr extractions using the ADP method.
(Since the ADP method is not capable of effectively analyzing the
$L$ peak, only $K$ peak results can be presented.)  The result of
the global fit to the eight extractions is
11.2~$^{+1.8}_{-1.7}$/day, in good agreement with the $K$ peak
result that used the waveform measurement of rise time to select
events.

     Extraction 1 had a slight counting anomaly.  The waveform
digitizer was inoperative for the first 2.6 days of counting and
only ADP information was available.  During this short time
period the events selected by ADP in the $K$ peak were used to
supplement those chosen by waveform analysis and no selection of
$L$-peak events was made.

     As described above, four of the eight extractions were
followed with a second extraction.  Three of these (Cr 2-2, Cr
4-2, and Cr 6-2) were counted in a similar way to the primary
extractions, and the results are given in Table \ref{ADP results}
(Cr 2-2 is missing because the counter failed).  Because of the
limited number of data acquisition channels which included a
digitizer, these extractions were counted in an electronic system
that was only able to make the ADP measurement of pulse shape.
The combined results of extractions Cr 4-2 and Cr 6-2 showed no
\nuc{71}{Ge} production by the \nuc{51}{Cr} source, as expected.

\section{Systematic Effects in the Measurement of the Production
Rate}

\subsection{Uncertainty in overall efficiency}

\begin{table*}[ht]
\caption{Summary of the contributions to the systematic
uncertainty in the measured neutrino capture rate.  Unless
otherwise stated, all uncertainties are symmetric.  The total is
taken to be the quadratic sum of the individual contributions.
For comparison, many of the systematics for the solar neutrino
extractions are also provided.  (Some of the solar values depend
critically on the particular data set considered and are thus
missing.)  The statistical uncertainty in the result of the Cr
experiment is $^{+11.1}_{-10.5}$ \%.}
\label{systematic uncertainties}
\begin{tabular}{l d d}
                                                          &
\multicolumn{2}{c}{Uncertainty in percent}          \\ \cline{2-3}
Origin of uncertainty                                     &            for
solar runs                                                &         for Cr
runs                                                      \\
\hline
Chemical extraction efficiency                                            
                                                                        \\
\hspace*{10mm} Mass of added Ge carrier                   &              2.1
&                                                        2.1            \\
\hspace*{10mm} Amount of Ge extracted                     &              2.5
&                                                        3.5            \\
\hspace*{10mm} Carrier carryover                          &              0.5
&                                                        0.5            \\
\hspace*{10mm} Mass of gallium                            &              0.5
&                                                        0.5            \\
\hspace*{5mm} Chemical extraction subtotal                &              3.3
&                                                        4.1            \\
\\
Saturation factor                                                         
                                                                        \\
\hspace*{10mm} Exposure time                              &              0.14
&                                                         0.            \\
\hspace*{10mm} Lead time                                  &              0.8
&                                                         0.            \\
\hspace*{5mm} Saturation factor subtotal                  &              0.8
&                                                         0.            \\
\\
Counting efficiency                                       &               
&                                                                       \\
\hspace*{10mm} Calculated efficiency                      &               
&                                                                       \\
\hspace*{15mm} Volume efficiency                          &              0.5
&                                                        0.5            \\
\hspace*{15mm} Peak efficiency                            &              2.5
&                                                        2.5            \\
\hspace*{15mm} Simulations to correct for counter filling &              1.7
&                                                        1.7            \\
\hspace*{10mm} Calibration statistics                     &               
&                                                                       \\
\hspace*{15mm} Centroid                                   &              0.1
&                                                        0.1            \\
\hspace*{15mm} Resolution                                 &              0.3
&                                                        0.3            \\
\hspace*{15mm} Rise time cut                              &              0.6
&                                                        0.6            \\
\hspace*{10mm} Gain variations                            &            ---
&                                                       $+$2.0          \\
\hspace*{10mm} Rise time window offset                    &            ---
&                                                         0.            \\
\hspace*{5mm} Counting efficiency subtotal                &
$+$4.4,$-$3.2                                             &
$+$3.7,$-$3.1                                             \\
\\
Residual radon after time cuts                            &            ---
&                                                       $-$1.7          \\
Solar neutrino background                                 &              0.
&                                                        1.2            \\
\nuc{71}{Ge} carryover                                    &              0.
&                                                        0.3            \\
\\
Total systematic uncertainty                              &            ---
&                                                   $+$5.7,$-$5.6
\end{tabular}
\end{table*}

     A summary of the various contributions to the overall
systematic uncertainty is given in Table \ref{systematic
uncertainties}.  Most of these components are the same as for the
solar neutrino extractions; so the values for the solar runs are
also given in Table \ref{systematic uncertainties} for
comparison.  The overall efficiency is the product of three
factors: the chemical extraction efficiency, the saturation
factor, and the counting efficiency.  The uncertainty in each of
these efficiencies will now be considered.

     The major components of the uncertainty in the chemical
extraction efficiency were the amount of Ge carrier added, the
measured amount of Ge carrier extracted, and the amount of
residual Ge carrier remaining from previous extractions.  The
concentration of Ge in the Ga:Ge alloy that was added as carrier
was measured by atomic absorption spectroscopy and isotope
dilution spectroscopy.  The resultant total uncertainty in the
amount of carrier added was \+/-2.1\%.  There was a \+/-3.5\%
uncertainty in the measurements of the amount of Ge that was
extracted; this value was larger than for the solar runs because
of the smaller number of extractions performed.  There were also
\+/-0.5\% uncertainties in the amount of Ga and the amount of
residual Ge carrier.  Adding these components in quadrature
yields a total uncertainty in the chemical extraction efficiency
of \+/-4.1\%.

     The saturation factor for the Cr source [for solar
neutrinos] is defined as the factors that multiply
$p_{\text{Cr}} \epsilon^k$ [$p_\odot \epsilon^k$] on the
right-hand side of Eq.~(\ref{aCr}) [Eq.~(\ref{asolar})].  The
time of exposure to the source $\theta_{\text{Cr}}$ depended only
on when the source was inserted and removed from the
Ga-containing reactor, and was very well established; so the
uncertainty in the source saturation factor was negligible.
There was a minor uncertainty in the time of solar exposure
$\theta_\odot$ because the extraction was made from two reactors
and the mean time of extraction was used as the end time of
exposure.  But since the production rate from solar neutrinos was
much less than from the Cr source, the uncertainty in the solar
saturation factor was also negligible.

     The uncertainty in the calculated counting efficiency was
mentioned in Sec.\ \ref{counting}.  The three components are the
uncertainty in volume efficiency (0.5\%), in measurements to
determine peak efficiency (2.5\%), and in simulations used to
correct for differing GeH$_4$ percentages and counter pressures
(1.7\%), giving a combined uncertainty of 3.1\%.  The uncertainty
in counting also includes the statistical uncertainty arising
from the limited number of events in the \nuc{55}{Fe}
calibrations, which typically had 1000--5000 events each.  There
were \+/-0.1\%, \+/-0.3\%, and \+/-0.6\% uncertainties in the
counting efficiency due to the uncertainties in the extrapolated
\nuc{71}{Ge} $L$- and $K$ peak centroid, resolution and rise time
limits, respectively.  Finally, there was a +2.0\% uncertainty
due to gain variations during the time that the \nuc{71}{Ge} was
decaying.  This value is one sided because gain drifts can only
shift the \nuc{71}{Ge} peak out of the event selection window.
Adding these uncertainties in quadrature gave a total uncertainty
in the production rate of $^{+3.7}_{-3.1}$ \% due to the
uncertainty in the counting efficiency.

\subsection{Other systematic uncertainties}

     The final uncertainty that is common to both the Cr
experiment and the solar neutrino measurements arises from the
inefficiency of a 3.25-h time cut for Rn that might be added to
the counter at the time it was filled.  By analyzing the first
five extractions both with and without this cut, we found that it
removed a total of 22 events assigned to \nuc{71}{Ge}.  Since the
cut deletes all but 10\% of false \nuc{71}{Ge} events, this
implies that 2.2 false events may remain after the cut.  As the
total number of events assigned to \nuc{71}{Ge} in these five
runs after the cut was 129.4, the systematic uncertainty after
the cut was thus $-1.7$\%.  The value is negative since radon
decays mistakenly identified as \nuc{71}{Ge} can only increase
the observed signal.

     Two systematic errors in Table \ref{systematic
uncertainties} are unique to the Cr-source experiment.  As
discussed in Sec.\ \ref{maximum likelihood analysis}, there is an
additional contribution to the measured signal from solar
neutrinos and there is a carryover correction due to the
incomplete removal of \nuc{71}{Ge} in the previous chemical
extraction.

     Although the production of \nuc{71}{Ge} in 13 tonnes of Ga
by solar neutrinos is small, it is finite and a correction is
necessary.  We took the solar neutrino capture rate to be 69 SNU
\cite{Abdurashitov et al. 94} and subtracted from the observed
signal an amount corresponding to that production rate.  The
solar neutrino rate has been measured by SAGE to a precision of
12 SNU or 17\%.  However, the solar neutrino production was only
a 6.8\% correction (9.8 events out of 143.7) and thus its
uncertainty resulted in a small (1.2\%) uncertainty in the
measured \nuc{51}{Cr} production rate.

     The efficiency for extracting Ge from the Ga was typically
85\%.  Thus a fair amount of Ge remained in the Ga after
extraction.  Immediately following extractions for the solar
neutrino runs, a second extraction is usually carried out.
Because of these second extractions and because the time between
extractions is several \nuc{71}{Ge} lifetimes, the number of Ge
atoms that survive to the end of the next solar run is
negligible.  In the Cr experiment extraction schedule, however,
this was not always the case.  Second extractions were conducted
only after extractions 2, 4, 6, and 7; so extractions 2, 4, and 6
contain a small contribution from \nuc{71}{Ge} produced during
the previous exposure.  The total number of events ascribed to
\nuc{71}{Ge} was 143.7 with an uncertainty of approximately 10\%.
The total number of estimated carryover events was 4.0 which is
determined with the same 10\% uncertainty.  Therefore the
uncertainty in the \nuc{51}{Cr} production rate due to the
uncertainty in the carryover correction was 0.3\%.

\begin{table*}[ht]
\caption{Values and uncertainties of the terms that enter the
calculation of the cross section.  All uncertainties are
symmetric.}

\label{expected rate uncertainties}
\begin{tabular}{l d d d}
                                                                                                               &
                                                                                                               &
\multicolumn{2}{c}{Uncertainty}                                                                          \\ \cline{3-4}
Term                                                                                                           &
Value                                                                                                          &
Magnitude                                                                                                      &
Percentage                                                                                                     \\
\hline
Atomic density $D = \rho N_0 f_I/M$                                                                            &
                                                                                                               &
                                                                                                               &
                                                                                                               \\
\hspace*{10mm} Ga density $\rho$ (g Ga/cm$^3$) \cite{Koster 70}                                                &
6.095                                                                                                          &
0.002                                                                                                          &
0.033                                                                                                          \\
\hspace*{10mm} Avogadro's number $N_o$ ($10^{23}$ atoms Ga/mol)                                                &
6.0220                                                                                                         &
negligible                                                                                                     &
negligible                                                                                                     \\
\hspace*{10mm} \nuc{71}{Ga} isotopic abundance $f_I$ (atoms
\nuc{71}{Ga}/100 atoms Ga)\cite{Machlan et al. 86}                                                             &
39.8921                                                                                                        &
0.0062                                                                                                         &
0.016                                                                                                          \\
\hspace*{10mm} Ga molecular weight $M$ (g Ga/mol) \cite{Machlan
et al. 86}                                                                                                     &
69.72307                                                                                                       &
0.00013                                                                                                        &
0.0002                                                                                                         \\
\hspace*{5mm} Atomic density $D$ ($10^{22}$ atoms
\nuc{71}{Ga}/cm$^3$)                                                                                           &
2.1001                                                                                                         &
0.0008                                                                                                         &
0.037                                                                                                          \\
Source activity at reference time $A$ ($10^{16}$ \nuc{51}{Cr}
decays/s)                                                                                                      &
1.9114                                                                                                         &
0.0022                                                                                                         &
1.2                                                                                                            \\
Capture rate $p$ (\nuc{71}{Ge} atoms produced/day) (uncertainties
combined in quadrature.)                                                                                       &
14.0                                                                                                           &
1.7                                                                                                            &
12.1                                                                                                           \\
Path length in Ga $\langle L \rangle$ (cm)                                                                     &
72.6                                                                                                           &
0.2                                                                                                            &
0.28                                                                                                           \\
\\
Cross section $\sigma$ [$10^{-45}$ cm$^2$/(\nuc{71}{Ga} atom
\nuc{51}{Cr} decay)]                                                                                           &
5.55                                                                                                           &
0.68                                                                                                           &
12.3
\end{tabular}
\end{table*}

\section{Measured Production Rate}
\label{production rate}

     The quadratic combination of all the systematic
uncertainties described in the last section is
$^{+5.7}_{-5.6}$\%.  The measured production rate in the $K$ and
$L$ peaks given in Section \ref{results}, including both
statistical and systematic errors, thus becomes
$p_{\text{Cr}}$~=~14.0 \+/-~1.5~(stat) \+/-~0.8~(syst) atoms of
\nuc{71}{Ge} produced per day.  This production rate is
equivalent to about 3500 SNU, 50 times higher than the rate from
solar neutrinos.

     For comparison, in the GALLEX \nuc{51}{Cr} experiments
\cite{Hampel et al. 98}, the average measured source production
rate at the beginning of the first exposure was 11.1 \nuc{71}{Ge}
atoms per day and the production rate from solar neutrinos and
other background sources was 0.7/d.  Even though our source had
one-third the intensity of a GALLEX source, our production rate
was nearly one-third higher and our background rate (see Sec.\
\ref{maximum likelihood analysis}) was a factor of 3 lower.  This
illustrates the significant advantage of using Ga metal with its
high atomic density as the target for a neutrino source
experiment.  Further, our source had very high enrichment and
consequent small physical size, leading to a long path length
through the gallium absorber.

\section{Measured Neutrino Capture Cross Section}

     For a neutrino source of activity $A$, it follows from the
definition of the cross section $\sigma$ that the capture rate
$p$ of neutrinos in a material around the source can be written
as the product
     \begin{equation}
     \label{cross section}
     p = A D \langle L \rangle \sigma,
     \end{equation}
\noindent where $D = \rho N_0 f_I/M$ is the atomic density of the
target isotope (see Table \ref{expected rate uncertainties} for
the values and uncertainties of the constants that enter $D$),
and $\langle L \rangle$ is the average neutrino path length
through the absorbing material, which in the case of a
homogeneous source that emits isotropically is given by
     \begin{equation}
     \langle L \rangle = \frac{1}{4 \pi V_S}
                         \int_{\text{absorber}} dV_A
                         \int_{\text{source}}
                         \frac{dV_S}{r^{2}_{SA}}.
     \end{equation}
\noindent In this last equation $r_{SA}$ is the distance from
point $S$ in the source to point $A$ in the absorber and the
source and absorber volumes are $V_S$ and $V_A$, respectively.

     The Ga-containing reactor in which the \nuc{51}{Cr} source
was placed was nearly cylindrical, with a dished bottom.  Based
on accurate measurements of the reactor shape, the path length
$\langle L \rangle$ was determined by Monte Carlo integration
over the source and absorber volumes to be 72.6~\+/- 0.2 cm.  The
accuracy of this integration was verified by checking its
predictions for geometries that could be calculated analytically
and by noting that the measured Ga mass contained in the reactor
volume agreed with that predicted by the integration.  The
sensitivity of $\langle L \rangle$ to the reactor geometry, to
the position of the source in the Ga, and to the spatial
distribution of the source activity were all investigated by
Monte Carlo integration, and the uncertainty given above includes
these effects.

     Substituting our measured values of $p_{\text{Cr}}$
(Sec.\ \ref{production rate}) and $A$
(Sec.\ \ref{source activity determination}), and the constants
$D$ (Table~\ref{expected rate uncertainties}) and $\langle L
\rangle$ into Eq.~(\ref{cross section}), we obtain
     \begin{eqnarray}
     \label{cross section result}
     \sigma & = & [5.55
               \pm 0.60 \text{ (stat)}\pm 0.32\text{ (syst)}]\\
               \nonumber
               & &\times 10^{-45}
               \frac{\text{cm}^2}
               {\mnuc{71}{Ga} \text{ atom }
                \mnuc{51}{Cr} \text{ decay}}.
     \end{eqnarray}
\noindent Because the half life for the \nuc{71}{Ge} to
\nuc{71}{Ga} decay is well known, the part of this cross section
that is due to the transition to the \nuc{71}{Ge} ground state
can be accurately calculated.  The value given by Bahcall
\cite{Bahcall 97} is 5.53 \E{-45} cm$^2$.  The portion of our
experimentally determined cross section that can result from
transitions to the two other states in \nuc{71}{Ge} which can be
excited by \nuc{51}{Cr} neutrinos (at 175 keV and 500 keV above
the \nuc{71}{Ge} ground state) is thus (0.02 \+/- 0.68) \E{-45}
cm$^2$.  Alternatively, as shown by Hata and Haxton \cite{Hata
and Haxton 95}, by taking the ratio of the measured cross section
to the ground state cross section, our measurement restricts the
weak interaction strengths BGT of these two levels according to
\begin{eqnarray}
     1 & + & 0.667\frac{\text{BGT(175 keV)}}{\text{BGT(g.s.)}} \\
                                                       \nonumber
       & + & 0.218 \frac{\text{BGT(500 keV)}}{\text{BGT(g.s.)}}
     = 1.00 \pm 0.12,
\end{eqnarray}
\noindent where BGT(g.s.) = 0.087 \+/- 0.001 is the strength of
the transition to the \nuc{71}{Ge} ground state.

\section{Discussion}

     The primary motivation for the \nuc{51}{Cr} source
experiment was to determine if there is any unexpected problem in
either the chemistry of extraction or the counting of
\nuc{71}{Ge}, i.e., to see if there is some unknown systematic
error in one or both of the efficiency factors in $\epsilon$, the
product of extraction and counting efficiencies.  If some such
systematic error were to exist, then the value of $\epsilon$ that
we have used in the preceding will be in error by the factor $E$,
defined as $E \equiv \epsilon_{\text{true}}/
\epsilon_{\text{measured}}$.  Since the cross section is
inversely proportional to $\epsilon$, this hypothetical error is
equivalent to the cross section ratio,
$E = \sigma_{\text{measured}}/\sigma_{\text{true}}$.  An
experimental value for $E$ can be set from our measured cross
section, Eq.~(\ref{cross section result}), if one assumes that
the true cross section is equivalent to the theoretically
calculated cross section.  Then $E \approx R \equiv
\sigma_{\text{measured}}/\sigma_{\text{theoretical}}$.  Neutrino
capture cross sections averaged over the four neutrino lines of
\nuc{51}{Cr} have been calculated by Bahcall \cite{Bahcall 97}
and by Haxton \cite{Haxton 98}.

     Bahcall, assuming that the strength of the two excited
states in \nuc{71}{Ge} that can be reached by \nuc{51}{Cr}
neutrinos is accurately determined by forward-angle $(p,n)$
scattering, gives a result of 5.81 (1.0 $^{+0.036}_{-0.028})$
\E{-45} cm$^2$.  The upper limit for the uncertainty was set by
assuming that the excited state strength could be in error by as
much as a factor of 2; minor contributions to the uncertainty
arise from forbidden corrections, the \nuc{71}{Ge} lifetime, and
the threshold energy.

     An independent consideration of the contribution of excited
states has been made by Hata and Haxton \cite{Hata and Haxton 95}
and very recently by Haxton \cite{Haxton 98}.  They argue that,
because of destructive interference between weak spin and strong
spin-tensor amplitudes in \nuc{71}{Ge}, the strengths determined
from $(p,n)$ reactions are, for some nuclear levels, poor guides
to the true weak interaction strength.  In particular, Haxton
finds the weak interaction strength of the $(5/2)^-$ level in
\nuc{71}{Ge} at an excitation energy of 175 keV to be much
greater than the value that is measured by the $(p,n)$ scattering
reaction, and calculates a total \nuc{51}{Cr} cross section of
(6.39~\+/- 0.68) \E{-45} cm$^2$.  This cross section was deduced
from the measured $(p,n)$ cross sections for the two excited
states, and uses a large-basis shell model calculation to correct
for the presence of spin-tensor contributions.  Since not all
known theoretical uncertainties were included, the stated error
here is a lower bound.

     Combining our statistical and systematic uncertainties for
the cross section in quadrature into an experimental uncertainty,
we can thus give estimates for $E$:
     \begin{eqnarray}
E & \approx & R \equiv \frac{\sigma_{\text{measured}}}
       {\sigma_{\text{theoretical}}} \\
& = & \left\{ \begin{array}{l l}
0.95 \pm 0.12 \text{ (expt)}\ ^{+0.035}_{-0.027} \text{ (theor)}
     & \text{ (Bahcall)}, \\
0.87 \pm 0.11 \text{ (expt)}\ \pm 0.09 \text{ (theor)}
     & \text{ (Haxton)}.
     \end{array}
     \right.
     \nonumber
     \end{eqnarray}
\noindent With either of these theoretical cross sections, $R$ is
consistent with unity, which implies that the total efficiency of
the SAGE experiment to the neutrinos from \nuc{51}{Cr} is close
to 100\%.

     The measurement reported here should not be interpreted as a
direct calibration of the SAGE detector for solar neutrinos.
This is because the \nuc{51}{Cr} neutrino spectrum differs from
the solar spectrum, there is a 10\%--15\% uncertainty in the
theoretical value for the \nuc{51}{Cr} cross section, and the
total experimental efficiency for each solar neutrino measurement
is known to a higher precision than the 12\% experimental
uncertainty obtained with the \nuc{51}{Cr} source.  As a result,
the solar neutrino measurements reported by SAGE should not be
scaled by the factor $E$.  Rather, we consider the Cr experiment
as a test of the experimental procedures, and conclude that it
has demonstrated {\it with neutrinos} that there is no unknown
systematic uncertainty at the 10\%--15\% level.

     The neutrino spectrum from \nuc{51}{Cr} is very similar to
that of \nuc{7}{Be}, but at slightly lower energy.  Since the
response of \nuc{71}{Ga} to \nuc{7}{Be} neutrinos is governed by
the same transitions that are involved in the \nuc{51}{Cr} source
experiment, we can definitely claim that, if the interaction
strength derived from the \nuc{51}{Cr} experiment is used in the
analysis of the solar neutrino results, then the capture rate
measured by SAGE includes the full contribution of neutrinos from
\nuc{7}{Be}.  This observation holds independent of the value of
$E$ or of cross section uncertainties.  This demonstration is of
considerable importance because a large suppression of the
\nuc{7}{Be} neutrino flux from the sun is one consequence of the
combined analysis of the four operating solar neutrino
experiments \cite{Bludman et al. 93,Akhmedov et al. 95}.

     GALLEX has completed two \nuc{51}{Cr} measurements whose
combined result, using the cross section of Bahcall \cite{Bahcall
97}, can be expressed as $R$ = 0.93~\+/- 0.08 \cite{Hampel et al.
98}, where the uncertainty in the theoretical cross section has
been neglected.  Both SAGE and GALLEX, which employ very
different chemistries, give similar results for the solar
neutrino capture rate and have tested their efficiencies with
neutrino source experiments.  The solar neutrino capture rate
measured in Ga is in striking disagreement with standard solar
model predictions and there is considerable evidence that this
disagreement is not an experimental artifact.

\section*{Acknowledgments}

     We thank E.~N.~Alexeyev, J.~N.~Bahcall, M.~Baldo-Ceolin,
L.~B.~Bezrukov, S.~Brice, A.~E.~Chudakov, G.~T.~Garvey,
W.~Haxton, P.~M.~Ivanov, H.~A.~Kurdanov, V.~A.~Kuzmin,
V.~V.~Kuzminov, V.~A.~Matveev, R.~G.~H.~Robertson, V.~A.~Rubakov,
L.~D.~Ryabev, and A.~N.~Tavkhelidze for stimulating our interest
and for fruitful discussions.  We acknowledge the support of the
Russian Academy of Sciences, the Institute for Nuclear Research
of the Russian Academy of Sciences, the Russian Ministry of
Science and Technology, the Russian Foundation of Fundamental
Research, the Division of Nuclear Physics of the U.S. Department
of Energy, and the U.S. National Science Foundation.  This
research was made possible in part by grant No.\ M7F000 from the
International Science Foundation, grant No.\ M7F300 from the
International Science Foundation and the Russian Government, and
award No.\ RP2-159 by the U.S. Civilian Research and Development
Foundation.

\end{document}